\documentclass[preprint2]{aastex}

\shorttitle{OBSERVATIONS OF LYMAN-ALPHA ABSORPTION IN A TRIPLE QUASAR SYSTEM}
\usepackage{subfigure}
\usepackage{graphics}
\begin{document}

\title{OBSERVATIONS\footnote{Based on observations with the NASA/ESA Hubble Space Telescope, obtained at the Space Telescope Science Institute, which is operated by the Association of Universities for Research in Astronomy, Inc., under NASA contract NAS5-26555.} \ OF LYMAN-ALPHA ABSORPTION IN A TRIPLE QUASAR SYSTEM}
\author{P. A. Young and C. D. Impey}
\affil{Steward Observatory, The University of Arizona, 933 N. Cherry Av. Tucson AZ 85721}\email{payoung@as.arizona.edu, cimpey@as.arizona.edu}
\and
\author{C. B. Foltz}
\affil{Multiple Mirror Telescope Observatory, The University of Arizona, 933 N. Cherry Av. Tucson, AZ 85721}\email{cfoltz@as.arizona.edu}

\begin{abstract}

We have obtained follow-up observations of the quasar pair LBQS 0107$-$025A,B and new observations of the nearby quasar LBQS 0107$-$0232 with the Hubble Space Telescope Faint Object Spectrograph. Extended wavelength coverage of LBQS 0107$-$025A and B using the G270H grating was also obtained. This triple system is unique in providing sensitivity to coherent Ly$\alpha$ absorption on transverse scales of approximately 1 Mpc at $z <$ 1. Monte Carlo simulations were used to establish the confidence level for matches between absorption features in different lines of sight as a function of velocity separation. Pairwise, there are 8, 9, and 12 lines that match between spectra. Three instances of matches between all three lines of sight were found with velocity separations of less than 550 $\rm{km\ s^{-1}}$. Two of the pairings have coincident lines within $\rm{|\Delta v| < 200 kms^{-1}}$ that would occur with less than 10\%\ probability by chance. Taking into account the equivalent widths of the lines, one of these triple coincidences is significant at the 99.99\% confidence level, based on Monte Carlo simulations with random line placements. Matches with strong lines preferentially have small velocity separations. These same simulations are used to demonstrate that the distribution of matches for a population of absorbers randomly distributed in velocity space is peaked toward $|\Delta$v$|$ = 0, which has implications for the statistical significance of matches. One of the triple coincidences appears to be a strong absorber with a sheet-like, but inhomogeneous, geometry and a coherence length approaching or exceeding 1 Mpc.

\end{abstract}

\keywords{quasar: absorption lines - intergalactic medium - cosmology: observations}

\section{INTRODUCTION}

Quasar absorption lines offer a remarkable opportunity to study the large scale structure of the universe. Given a sufficiently bright background illuminating source and a combination of observations from the ground and space, their properties can be studied over 90-95\% of the Hubble time. Much recent progress has been made using the unique spectroscopic capabilities of the Hubble and Keck telescopes. Interest in low column density hydrogen absorbers -- the Lyman $\alpha$ forest -- is also motivated by theoretical insights. The absorbers are valuable cosmological tools because (a) they are as common as luminous galaxies, (b) they contain most of the baryons in the universe, (c) they accurately trace the underlying dark matter potential, and (d) at low column density they may be primitive enough to retain a memory of initial conditions, unlike highly non-linear objects such as galaxies \citep{cen94, zha95, her96, mir96, dav99}. The fact that supercomputer simulation can reproduce the basic features of the Lyman $\alpha$ forest has become an in important argument in support of the standard picture of hierarchical structure formation. 

The statistical properties of quasar absorbers are usually deduced from the combination of observations along single, widely separated sight-lines. At high redshift, the HIRES spectrograph of the Keck telescope has provided the sensitivity and resolution needed to resolve the Lyman $\alpha$ forest \citep{hue95, kim97, kir97}. Below a redshift of $z = 1.6$, we must rely on the modest aperture and moderate resolution of the spectrographs aboard the Hubble Space Telescope. The demographics of the absorbers have been defined by the Quasar Absorption Line Key Project \citep{bah96, jan98, wey98}. Quasar pairs or groups that are closely aligned in the sky can be used to measure the transverse size or the coherence or clustering scale of the absorbers. The experiment is conceptually simple: if Lyman $\alpha$ absorption is seen at a similar velocity in two lines of sight and the chance probability of such a line match is small, then the transverse separation is a firm lower limit on the size of the cloud. Guided by the insights of supercomputer simulations, it is clear that the idea of the absorber as ``clouds'' with any simple geometry is naive. Even the notion of a characteristic size or radius is imprecise -- the absorbers trace an extensive and complex topology.  We use the term 'size' throughout this study but we realize that it could refer to the physical size of a single absorbing entity or the coherence length of an ensemble of smaller absorbing clouds.

Until a few years ago there were few size constraints on the hydrogen clouds. Common Lyman $\alpha$ absorption in quasars that were gravitationally lensed had been used to derive lower limits to the sizes of a few tens of kpc \citep{fol84, sme92}. The first evidence of large absorber sizes (hundreds of kpc) came from studies of the quasar pair LBQS 1343$+$264A,B \citep{din94, bec94}. The characteristic size deduced from these studies, $\sim$ 300 $h_{70}^{-1}$ kpc at $z \sim 2$ ($q_0 = 0.5$), rules out simple absorber models such as cold dark matter mini-halos \citep{ree86} and pressure-confined clouds \citep{sar80}. Since then, a number of pairs have been observed to extend these experiments to larger transverse separations and lower redshifts \citep{din95, elo95, fan96, cro98, din98, pet98}. These experiments are still limited by the modest number of lines available for any analysis of coincidence. it is still difficult to distinguish between models where the absorbers are (a) non-spherical, or (b) clustered, or (c) spherical but drawn from a non-uniform size distribution. In all these analyses the addition of a third sight-line adds important information about the two-dimensional structure of absorbers in the plane of the sky.

The next obvious step is to relate observations of the Lyman-$\alpha$ absorbers to the results from the supercomputer simulations. This is non-trivial --- there are major differences in methodology between the ways observers and modelers measure absorber properties \citep{mir96, cha97, dav99}. Observers have one dimensional path-lengths but cannot sample one contiguous volume even with multiple sight-lines. Supercomputer simulations can be used to measure large scale structures on all scales up to the maximum size of the experiment, about 15-40 $h_{70}^{-1}$ Mpc. Spectral extractions from the simulations have high spectral resolution, infinite signal-to-noise and a known continuum level, while observers must deal with photon statistics and the vagaries of continuum-fitting and line identification. Modelers know the physical state of the gas; observers must make operational definitions for the detection, profile-fitting, and de-blending of spectral features. Our long term goal is to make direct comparisons with the results of N-body simulations that incorporate gas dynamics \citep{kat96}. 

Although we defer a detailed interpretation of the absorber coincidences until a future paper, we make several improvements to the methodology of defining coincident lines herein. First, we impose no velocity window on the definition of a line match in an adjacent sight-line, in contrast to virtually every published paper in this subject. The rationale is that coincident line pairs should be selected without any {\it a priori} kinematic assumption about the absorber. We also include cosmic variance in line counting in the determination of the errors associated with line matching, which has also been rarely done in the past. Finally, we employ a full Monte Carlo simulation of a random line matching experiment as a way of deriving reliable significance levels to the results presented here. 

This paper presents new observations of the quasar pair LBQS 0107$-$025A,B, along with observations of a third quasar, LBQS 0107$-$0232, which creates baselines with up to 2.5 times the transverse separation of the original pair. The next section discusses the HST/FOS observations and data reduction procedure. We then proceed to describe the line matching procedure and the observed matches along with their significance as determined by comparison with Monte Carlo simulations of random populations of absorbers. The last section of the paper summarizes the implications of these observations for the geometry of the absorbers. The details of the Monte Carlo procedure are presented in Appendix A.

\section{OBSERVATIONS}

The program QSOs are listed in Table 1 with their J2000 coordinates, V magnitudes, and emission redshifts. The pair LBQS 0107$-$0232,25A has a LOS separation of 2\farcm 95. The pair LBQS 0107$-$232,25B has a LOS separation of 1\farcm 94. The final pair, LBQS 0107$-$025A,B, has a LOS separation of 1\farcm 29. For ${H_0}$ = 70 $\rm{km\ s^{-1}\ Mpc^{-1}}$ and ${q_0}$ = 0.5, these translate into projected separations of 990, 650, and 430 kpc, respectively.
 
We obtained ultraviolet observations of the QSO pair LBQS 0107$-$025A,B on 1996 December 29 with the post-COSTAR Faint Object Spectrograph (FOS) on the Hubble Space Telescope (HST), using the red Digicon detector, a 1\arcsec\  circular aperture, and the G270H grating. This configuration gave a spectral resolution (FWHM) of roughly 2 \AA, or 220 $\rm{km\ s^{-1}}$ over the wavelength range from 2200-3200 \AA. One 25 minute and three 40 minute exposures were combined, resulting in a total of 145 minutes accumulated on A. One 27 minute and two 40 minute exposures were combined for B, resulting in a total accumulated time of 107 minutes. The signal to noise ratios per pixel (S/N) away from regions containing broad emission lines in the resulting combined spectra are approximately 23 for A and approximately 47 for B.  

We also obtained HST FOS observations of the neighboring quasar LBQS 0107$-$0232 on 1997 January 2 using the red Digicon detector in combination with the 1\arcsec\  circular aperture and the G190H grating. This configuration gave a spectral resolution of 1.4 \AA\  or 200 $\rm{km\ s^{-1}}$ over the wavelength range 1625-2300 \AA. One 25 minute and six 40 minute exposures were combined, resulting in a total of 265 minutes accumulated on LBQS 0107$-$0232. The S/N per pixel in the continuum is approximately 13 in regions away from broad emission lines. 

Finally, we re-observed LBQS 0107$-$025A on 1997 February 5 using the same settings used for LBQS 0107$-$0232 above. We combined five 40 minute and one 25 minute integrations, for a total accumulated time of 225 minutes. The S/N per pixel in the continuum reached was approximately 25. Despite the generally smooth operation of the HST science program, this proposal was subject to a byzantine series of misadventures that have extended a single set of observations over four cycles. As a result, we are still awaiting the last dataset for this project, to be taken with the STIS instrument.

The observations of LBQS 0107$-$025B in G190H were adopted from \citet{dinth97}. Those earlier observations were taken on 1994 February 12, using the same settings as for LBQS 0107$-$0232 above. A total time of 108 minutes was accumulated on B. The S/N per pixel exceeds 10 for wavelengths shortward of the emission feature at approximately 2000 \AA, and is approximately 22 for wavelengths longward of 2000\AA. This is not a sharp discontinuity. Comparison of the data for LBQS 0107$-$025A from the current observations and \citet{dinth97} reveal a wavelength offset of 1.72 \AA\ as a result of differences in the wavelength zero point determination. It is difficult to determine which zero point calibration is correct, however this does not affect the matching results. A wavelength shift of $-$1.72 \AA\ was applied to the LBQS 0107$-$025B data to register it with the reference frame of the current observations.

\subsection{The HST FOS Data and Calibrations}

The FOS spectra were reduced at STScI with the STSDAS pipeline processing facility. The data was flux calibrated and corrected for dead and disabled diodes. Filter grating wheel positioning errors were corrected using comparison PtCrNe lamp spectra obtained immediately before or after the science G190H exposures for LBQS 0107$-$025A and LBQS 0107$-$0232 \citep{dinth97}.

Target mis-centering on the aperture may cause a significant wavelength zero point error which is difficult to correct for. To minimize this effect, the same guide star was used for all three lines of sight, leading to an expected error of approximately 0.3\AA\ or 45 $\rm{km\ s^{-1}}$ \citep{dinth97}. We did not attempt to use the \ion{Al}{2} galactic line observed in LBQS 0107$-$0232 and LBQS 0107$-$025A as a zero point, because the line appears to be blended with Lyman gamma at $z$ = 0.72 and possibly \ion{Si}{2} at $z$ = 0.4. The spectra for the G190H observations are presented in Figures 1 and 2, and line lists in Tables 2, 3, and 4. All equivalent widths in the tables are observed equivalent widths. Lines 9 and 10 in LBQS 0107$-$025A are part of a blend and have large errors in equivalent width. Since they are part of a prominent feature in the spectrum they are included in Table 3 even though they fall below the 3$\sigma$ limit. They are not used in any part of the analysis.

In the G270H observations of LBQS 0107$-$025A and B, many Galactic metal lines (\ion{Mg}{1}, \ion{Mg}{2}, \ion{Fe}{2}, and \ion{Mn}{2}) were detected and used for the wavelength zero point calibration. It was assumed that Galactic absorptions take place at the same redshift along both LOS to null the systematic wavelength shift due to centering errors between the two LOS. The spectra for the G270H observations are presented in Figures 3 and 4, and the line lists are given in Tables 5 and 6.

The average total rms uncertainty in the wavelength solution with calibration spectra is 0.11 diodes, which is approximately 0.18 \AA\ or 27.5 $\rm{km\ s^{-1}}$ for the G190H filter. For the G270H filter without calibration spectra, the rms uncertainty is 0.16 diodes, or approximately 0.36 \AA\ or 40 $\rm{km\ s^{-1}}$ \citep{key97}. In both cases the rms uncertainty in the wavelength solution is considerably smaller than the velocity resolution of the data. Grating wheel non-repeatability can introduce an error of approximately 1 diode (1.63 \AA\ or 250 $\rm{km\ s^{-1}}$ in the G190H filter) between epochs, which may account for the wavelength zero point offset between the current observations and those in \citet{dinth97}.

\subsection{Line Selection and Identification}

Line selection and identification was performed using Cathy Petry's heavily modified version of Tom Aldcroft's ``findsl'' program \citep{ald93}. The program interactively and iteratively fits cubic splines to the continuum. Points deviating negatively by more than two standard deviations were rejected from subsequent iterations until the RMS of positive and negative residuals in the remaining points were equal. Where the resulting continuum was not well fit visually (i.e. near the broad emission lines in the spectra), the continuum was manually adjusted. For most parts of the spectra no intervention was necessary. 
 
De-blending, wavelength, and equivalent width determinations were performed by fitting unresolved Gaussian profiles to the lines. Although ideally the absorption lines should be fit with Voigt profiles \citep{car84}, the low column density of the absorbers, the small Doppler width of the individual Ly$\alpha$ lines, and the fairly low spectral resolution of the observations combine to make such a procedure unnecessary.

A master line list to be used in the matching between lines of sight for each member of the triplet was then generated. All lines that were identified as belonging to three metal systems previously identified at $z$ = 0.4, $z$ = 0.227, and $z$ = 0.289) by \citet{din95} were removed. A systematic search using a home-made program comparable to the Key Project ZSEARCH program \citep{sch93} revealed no new metal systems. The G190H spectra were then searched for high order Lyman lines. Two and four Ly$\beta$ lines were found in LBQS 0107$-$0232 and LBQS 0107$-$025A, respectively. These were all found to have counterpart Ly$\alpha$ lines at the same redshift and so were removed from the line list to avoid double counting coincidences. The final list of identified lines consisted of lines identified at the 3$\sigma$ level. The line lists used for matching were limited to 5$\sigma$. The wavelength range for the Ly$\alpha$ sample omits lines less than 3000 $\rm{km\ s^{-1}}$ from the emission line redshifts of the quasars in order to avoid the biases that might be introduced by the proximity effect. The positions of the lines used in matching are plotted side-by-side in Figure 5 for comparison.

The matching list for LBQS 0107$-$0232 contains 16 Ly$\alpha$ lines, 32 for LBQS 0107$-$025A, and 20 for LBQS 0107$-$025B. It should be noted that the sample is not complete down to a uniform equivalent width limit of either 0.21\AA, where the functional form of the distribution of lines with equivalent width changes form substantially, or 0.24\AA, which is the HST Key Project equivalent width limit. A plot of the equivalent width distribution in the full 3$\sigma$ data set (binned to 0.1 \AA\ and corrected to rest equivalent width) with the Key Project distribution for comparison is given in Figure 6. Four lines with a rest equivalent width above 1.0\AA\ are present in the sample, which is consistent with a prediction of $\approx$3.5 lines from the HST Key Project. The prediction of the number of lines uses $\gamma\ = 0.11\pm0.16$, $W^{*} = 0.272\pm0.008$, and $dN/dz = 35.6\pm3.2$ from the preferred sample in \citet{wey98}. 
 
\section{DATA ANALYSIS}

\subsection{Line Matching Procedure}

The nearest neighbor for each Ly$\alpha$ line in the other two spectra was found independently for each member of the triplet using the final line matching lists described in the previous section. As noted above, only lines with S/N$>$ 5 were used. No condition on velocity difference was imposed to make a match. The only requirement was that for a given line in LOS 1, the nearest neighbor in LOS 2 have the original line in LOS 1 identified as its nearest neighbor in order to be declared a match. This reversible matching criterion prevents multiple lines in one spectrum from being identified with a single line in the other and imposes a uniqueness criterion on the matches. It does, however, come at the cost of lines that might remain unmatched in either spectrum. 

The same matching procedure was extended to deal with matches across three lines of sight. For a given line in LOS 1 that is uniquely matched to lines in LOS 2 and 3, the lines in LOS 2 and 3 must also be uniquely matched to each other in order for the three lines to be counted as coincident across the triplet. The mean line separations in the three LOS are 2410, 2270, and 3580 $\rm{km\ s^{-1}}$ for LBQS 0107$-$0232, LBQS 0107$-$025A, and LBQS 0107$-$025B, respectively. We did not consider matches with $|\Delta$v$|$ more than half the mean line separations because such matches would be ambiguous due to severe aliasing. Velocity separations were calculated using the full relativistic expression for Doppler shifting, rather than the approximation $\Delta\lambda/\lambda\approx\Delta v/c$.

The choice of a velocity separation limit for discriminating between true coincidences and random matches has often been set rather arbitrarily, based upon either precedents in the literature or a visual examination of a histogram of velocity separations \citep{din97}. We prefer to impose no {\it a priori} limit on the kinematic properties of an absorber across these large scales. In order to assign a quantitative confidence level to our matches we compared our results to Monte Carlo simulations of randomly distributed populations of absorbers.

So far we have only discussed coincidences in terms of counting lines. This is misleading in that the varying signal to noise between the spectra must be taken into account. Therefore, the distribution of lines in wavelength is not homogeneous in each spectrum, and the limiting equivalent width also depends on wavelength. We addressed this issue by assigning equivalent widths to the lines in the Monte Carlo experiment. These are drawn from the equivalent width distribution for low redshift absorbers found by the HST Key Project \citep{wey98}, converted to observed equivalent width for the redshift of the line in question. Each line was randomly placed in the spectrum and was subjected to a selection procedure which models the variation in sensitivity with wavelength of the data. Details of the Monte Carlo simulations including this selection procedure are presented in Appendix A. The use of this procedure ensures that the simulations are carried out with lines that statistically mimic the properties of the observed lines.

\subsection{Double and Triple Coincident Lines}

All matches with $|\Delta$v$|$ less than one half the average velocity separation within the more sparsely populated line of sight are given in Tables 7-9. A histogram of the number of matches against $|\Delta$v$|$ is given in Figure 7, with the results of the corresponding Monte Carlo simulations superimposed. There are eight matches in the pair LBQS 0107$-$0232,025A, nine in LBQS 0107$-$0232,025B, twelve in LBQS 0107$-$025A,B, and three triple coincidences. Triple coincidences are summarized in Table 10 and Figure 7d. 

Note that most lines in each spectrum have no counterpart in either of the adjacent sight-lines. The fraction of unmatched lines in the first pairing above is 0.5 for LBQS 0107$-$0232 and 0.75 for LBQS 0107$-$025A. For the second pairing the fraction is 0.44 for LBQS 0107$-$0232 and 0.55 for LBQS 0107$-$025B. The fraction of unmatched lines for the third pairing is 0.63 for LBQS 0107$-$025A and 0.4 for LBQS 0107$-$025B. We also experimented with the effect on these numbers when the reversible match criterion was relaxed, so that a single line in one spectrum could be matched to more than one line in the adjacent sight-line. The corresponding fractions of unmatched lines when the uniqueness criterion was removed are 0.19 and 0.75 for the first pairing, 0.19 and 0.55 for the second pairing, and 0.63 and 0.3 for the third pairing. In all cases the non-unique matches were in the sight-line with the lower number density of absorbers.  

The probability that the cumulative number of observed matches with less than a given $|\Delta$v$|$ would exceed the number expected from the simulations was calculated for $|\Delta$v$|$$\ <$ 200, 400, and 600 $\rm{km\ s^{-1}}$. These values bracket those chosen to define matches in the literature. The results for the three velocity windows are given in Table 11. The probability is simply the fraction of trials in which the cumulative number of matches exceeds that seen in the data. The coincidences in the pairings LBQS 0107$-$0232,025B and LBQS 0107$-$025A,B show a smaller probability of being the product of random matches than those in LBQS 0107$-$0232,025A. However, in no case is the number of coincidences significant at more than the 91\%\ confidence level. The number of coincidences in the triplet is not significant.

The average observed equivalent widths of the lines in each pair as a function of the velocity separation are shown in Figure 8. The plot includes coincident lines from all three pairings. Matches with strong lines preferentially have small velocity separations. All of the coincidences with mean line strength above 1\AA\ have $|\Delta$v$|$$\ <$ 300 $\rm{km\ s^{-1}}$, whereas only 45\%\ (9/20) of the weaker coincidences have $|\Delta$v$|$$\ <$ 300 $\rm{km\ s^{-1}}$. This result supports the hypothesis that the coincidences among stronger lines are real, because the strong lines have a low chance coincidence rate within such a small velocity window.

We have detected coincident absorbers over $\approx$1 Mpc scales in two quasar pairs, using the velocity criterion $|\Delta$v$|$$\ <$ 200 $\rm{km\ s^{-1}}$ for a match. The weakness of the signal indicates either that the absorbers are not strongly clustered, or that only a fraction of the absorbers are coherent on these scales. It also demonstrates that excellent line-counting statistics are needed for a convincing detection of coherence. In fact, given the paucity of bright quasars with suitable separations, the targets do not exist to confirm or refute this result with current HST instrumentation.

The first triple coincidence listed in Table 10 is striking in terms of the large equivalent widths of the lines. A set of matches with these large equivalent widths did not occur a single time in the simulations among ten thousand trials. Therefore, inclusion of a consideration of line strength leads to the result that this triple coincidence is highly significant. We conclude that we have detected a spatially coherent absorber among all the three sight-lines of this experiment. However, we have no way of telling whether this single strong absorber is typical of the Ly$\alpha$ population.

Even if the coincident lines represent spatial coherence, the absorbers appear to be relatively inhomogeneous across these large transverse scales, as witnessed by the different equivalent widths of the lines in a match. A correlation test constrained to pass through the origin, performed on the EW vs. EW distribution of the matches in the data, returned a correlation coefficient of 0.38. (A similar test of the distributions in the simulations showed them to be uncorrelated, as expected.) For 29 matches this yields a 95\%\ chance that the equivalent widths in the data are correlated. Such a large scatter in column density from one sight-line to another cannot be used to argue among the rival explanations of clustered absorbers vs. inhomogeneities in a single large absorber.

\section{CONCLUSIONS}

We have presented new Hubble Space Telescope observations of the quasar pair LBQS 0107$-$025A,B, along with observations of a third quasar, LBQS 0107$-$0232. \citet{din97} found 5 coincident Lyman $\alpha$ lines in the quasar pair. The new data increase that yield to 12 coincident lines, albeit with a more generous velocity window for declaring a match. The comparisons of the LBQS 0107$-$025A,B pair to the third quasar yield 8 and 9 matches, respectively. There are 3 triple matches within a velocity window of 550 km s$^{-1}$.

The significance of the coincident absorbers is established with a Monte Carlo simulation that lays down lines randomly in velocity space, with numbers that account for the Poisson variance in the number of absorbers seen per sight-line and the varying signal to noise between spectra. Simulations of random lines give a distribution of coincident pairs that peaks slightly towards zero velocity difference, with important implications for determining the significance of matches in this (and other) experiments.  

The main results are, (1) coincident lines within $|\Delta$v$|$$\ <$ 200 $\rm{km\ s^{-1}}$ that are significant at the 90\%\ level for two quasar pairings, (2) coincident pairs of stronger lines preferentially have small velocity separations, and (3) a strong absorber that represents a highly significant triple coincidence.

The simplest interpretation of the double and triple coincidences is in terms of absorbers with coherence lengths spanning 0.5 to 1 Mpc at $0.4 < z < 0.9$ and extending in two dimensions on the plane of the sky. The correlation between line strengths for the coincident pairs is weak, so the absorbers must be extended but inhomogeneous. It is difficult to distinguish between single, coherent absorbers and clustered but distinct absorbers with a pair experiment. However, the detection of a highly significant match in all three quasars in the asterism indicates that at least one strong absorber has a sheet-like geometry. As more data on multiple quasar sight-lines accumulates, the line matching statistics can be used for an important new measure of the dimensionality of large scale structure at modest redshifts.

\acknowledgments

This research was supported by grant GO-06592.01-95A from NASA, awarded through the Space Telescope Science Institute. Development of some of the line-finding and identification software was supported by National Science Foundation grant AST 98-03072. We acknowledge the referee for helping improve the analysis and presentation of this paper. We thank Tom Aldcroft and Cathy Petry for making their spectral analysis packages available to us, and Christian Drouet d'Aubigny for substantial efforts with the data reduction. We appreciate the efforts of a number of STScI support staff to shepherd this long-running project through to near-completion.

\newpage

\appendix

\section{MONTE CARLO SIMULATIONS OF LINE MATCHING}

We simulate the matching of lines among two or three lines of sight, assuming that each quasar has a population of absorbers randomly distributed in velocity space. The number of absorbers in each line of sight was drawn from a Poisson deviate, the mean of which was taken to be the actual number of lines in the matching list for a given member of the triplet. This accounts for cosmic variance (with no additional large scale clustering) in the statistics of the absorbers. The same matching algorithm used on the data was applied to the simulated spectra. The result taken over a large number of trials gives the expected number of random matches for a given velocity separation. The simulation was performed for each of the three possible quasar pairings and for the three LOSs taken jointly. 

The behavior of the random distribution of matches is worthy of note. The number of possible combinations which can produce matches with a small velocity difference is greater than the number of combinations which can produce a large $|\Delta$v$|$. This produces a distribution which is naturally peaked toward $|\Delta$v$|$ = 0. To demonstrate this, consider two lines of sight binned to some fraction of the total velocity separation, each with one line randomly placed. The probability of a match for a given $|\Delta$v$|$ (in number of bins, with $L$ total bins in the LOS) is

\begin{displaymath}
P_1(\Delta v)=n P_{1,LOS1}(l) P_{1,LOS2}(m)
\end{displaymath}

where $P_{1,LOS1}(l)=f$ is the probability of a particular bin $l$ being occupied when there is one line and $n=2\times(L-\Delta v)$ is the number of combinations of line placement which can give a given $|\Delta$v$|$. Adding a second line in LOS 2 makes $P_{2,LOS2}(m)=1-(1-P_1(l))^2$. The probability of a match is changed by this factor and, for simple nearest neighbor matching, the probability of finding the second line at smaller $|\Delta$v$|$ than the first ($P_{small}=\Delta v/L$). The probability of finding a match for a given $|\Delta$v$|$ if a second line is added into LOS 1 is $P_2(\Delta v)=1-(1-P_1(\Delta v))^2$. Thus the total probability in a system with two lines in each line of sight is

\begin{displaymath}
P_2(\Delta v)=1-(1-nP_{1,LOS1}(l)P_{2,LOS2}(m)(1-P_{small}))^2
\end{displaymath}

This can be extended for larger numbers of lines in a similar fashion. The end result is a higher probability for small $|\Delta$v$|$ matches and a distribution peaking near zero. As the number of lines in each line of sight drops to one, the distribution approaches 

\begin{displaymath}
N = N_0 (0.5)^{\case{\Delta v}{0.25L}}
\end{displaymath}

The number of matches drops by one half approximately one quarter of the way through the velocity range of the sample. The distribution is more sharply peaked for large numbers of lines.

Heuristically, this can be argued as follows. There are more possible combinations which give velocity differences that are a small fraction of the total velocity range. Consider a situation where velocity separations are binned to one tenth of a total velocity range equal to unity. Only two combinations provide a $|\Delta$v$|$ of 1: velocities of 0 and 1 or 1 and 0. There are ten combinations which provide a $|\Delta$v$|$ of 0.5: 0 and 0.5, 0.1 and 0.6, and so on. Smaller separations will happen more often than large, and large ones will be further depleted by the probability of finding a match at smaller velocity when there is more than one line in each line of sight.

Simulations are presented here for pairs with 30 and 30, 30 and 20, and 30 and 10 lines in the two lines of sight with the same total range in velocity as the data to demonstrate the behavior of the distribution with unequal populations. The same procedure also shows how the cosmic variance in the line numbers might might affect the matching. The same simulations were also performed with the uniqueness criterion on matching removed. The matches were divided into $|\Delta$v$|$ bins of 25 $\rm{km\ s^{-1}}$. Careful centroiding of lines with velocity uncertainties comparable to those in the data ($\approx$ 200 $\rm{km\ s^{-1}}$) should be able to yield this accuracy. This data set is reliable to $\approx$ 200 $\rm{km\ s^{-1}}$, and the binning for the simulations which were used in the analysis reflects this. Ten thousand trials were performed for each simulation, and the total number of matches in each velocity bin after all the trials were completed was divided by the total number of trials. The results of these simulations are shown in Figure 9.

The sharpness of the peak is greater for equal numbers of lines than for the situation where the population of one LOS is reduced (i.e. 30 and 30 versus 30 and 10). A sample with unequal numbers of lines in the two sight lines is more sharply peaked than a symmetric sample with the smaller number of lines in each LOS (i.e. 30 and 10 versus 10 and 10). Applying the uniqueness criterion sharpens the peak by preferentially discarding higher $|\Delta$v$|$ matches to a given line.

This behavior has important implications for quasar absorption line studies. A larger number of matches at small velocity separations does not necessarily indicate a departure from a random population of absorbers. Appropriate simulations must be carried out in order to establish the significance of any particular number of coincidences. 

A further layer was added to the code which allows each line to be assigned an equivalent width. This was necessary to include the effects of differing signal to noise between two lines of sight in the simulation. A fourth order polynomial was fit to the rms errors in equivalent width to create an error function for each spectrum as a function of wavelength or, equivalently, $z$. For each line an equivalent width was drawn from the distribution found by the HST Quasar Absorption Line Key Project \citep{wey98}. This was corrected to the proper observed equivalent width for that line's redshift to make later comparison with the data more straightforward. This equivalent width was compared to the 5$\sigma$ sensitivity limit calculated from the error function appropriate to that simulated line of sight at that $z$. If the equivalent width was below the limit, the process was repeated until a line of sufficient strength was created. In essence, the simulation was overpopulated with lines, and all the lines which would have been too weak to be detected at the 5$\sigma$ level in the corresponding observed spectrum were discarded. The resulting simulations thus duplicate the signal to noise characteristics and population of the observed spectra and can be used as a direct test of the significance of matches.

\begin{deluxetable}{ccccc}
\footnotesize
\tablecaption{Program QSOs}
\label{tbl:table1}
\tablewidth{6in}
\tablewidth{0pt}
\tablehead{\colhead{QSO} & \colhead{R.A.} & \colhead{Dec.} & \colhead{V} & \colhead{$z_{\rm{em}}$} \\ \colhead{} & \colhead{(J2000)} & \colhead{(J2000)}}
\startdata
LBQS 0107$-$0232 & $\rm{01^{h}10^{m}}$14\fs 58 & $-$02\arcdeg 16\arcmin 57\farcs 2 & 18.4 & 0.728 \\
LBQS 0107$-$025A & 01 10 13.30 & $-$02 19 51.9 & 17.9 & 0.960 \\
LBQS 0107$-$025B & 01 10 16.35 & $-$02 18 50.8 & 17.3 & 0.956 \\
\enddata
\end{deluxetable}

\begin{deluxetable}{ccccccccc}
\footnotesize
\tablecaption{Absorption Lines in LBQS 0107$-$0232 (G190H)}
\label{tbl:table2}
\tablewidth{6in}
\tablewidth{0pt}
\tablehead{\colhead{No.} & \colhead{$\lambda$} &\colhead{$\sigma_{\lambda}$}
& \colhead{$W_{obs}$} & \colhead{$\sigma_{W}$} & \colhead{SNR} &\colhead{ID} &\colhead{$z$} & \colhead{comments} \\ \colhead{} & \colhead{(\AA)} & \colhead{(\AA)} & \colhead{(\AA)} & \colhead{(\AA)}}
\startdata
1 & 1702.12 & 0.30 & 0.84 & 0.19 & \phn4.4 & Ly$\alpha$ & 0.400 & Weak line sample \\
2 & 1753.68 & 0.21 & 0.66 & 0.14 & \phn4.9 & Ly$\beta$ & 0.710 & \\
3 & 1761.37 & 0.09 & 0.78 & 0.11 & \phn7.4 & Ly$\beta$ & 0.717 & \\
4 & 1782.21 & 0.24 & 0.56 & 0.10 & \phn5.7 & Ly$\alpha$ & 0.466 & \\
5 & 1853.27 & 0.09 & 0.87 & 0.10 & \phn8.4 & Ly$\alpha$ & 0.524 & \\
6 & 1857.31 & 0.10 & 0.85 & 0.14 & \phn6.1 & Ly$\alpha$ & 0.528 & \\
7 & 1864.19 & 0.18 & 2.48 & 0.20 & 12.6 & Ly$\alpha$ & 0.533 & \\
8 & 1877.99 & 0.09 & 1.35 & 0.12 & 11.4 & Ly$\alpha$ & 0.545 & \\
9 & 1892.33 & 0.07 & 3.63 & 0.15 & 24.1 & Ly$\alpha$ & 0.557 & Possible blend\\
10 & 1918.21 & 0.11 & 0.70 & 0.11 & \phn6.5 & Ly$\alpha$ & 0.578 & \\
11 & 1961.90 & 0.08 & 1.09 & 0.11 & \phn9.7 & Ly$\alpha$ & 0.614 & \\
12 & 1970.57 & 0.10 & 0.64 & 0.07 & \phn8.9 & Ly$\alpha$ & 0.621 & \\
13 & 2003.37 & 0.14 & 0.68 & 0.13 & \phn5.3 & Ly$\alpha$ & 0.648 & \\
14 & 2026.79 & 0.11 & 1.08 & 0.11 & \phn9.8 & Ly$\alpha$ & 0.667 & \\
15 & 2045.57 & 0.13 & 0.70 & 0.09 & \phn8.0 & Ly$\alpha$ & 0.683 & \\
16 & 2053.37 & 0.05 & 1.00 & 0.07 & 14.9 & Ly$\alpha$ & 0.689 & \\
17 & 2066.62 & 0.21 & 0.29 & 0.07 & \phn4.0 & Ly$\alpha$ & 0.700 & Weak line sample\\
18 & 2077.47 & 0.03 & 1.42 & 0.06 & 22.5 & Ly$\alpha$ & 0.709 & Blend with line 19\\
19 & 2079.21 & 0.09 & 0.42 & 0.04 & 10.2 & Ly$\alpha$ & 0.710 & \\
20 & 2086.59 & 0.13 & 0.50 & 0.16 & \phn3.1 & Ly$\alpha$ & 0.716 & Weak line sample\\
21 & 2087.81 & 0.09 & 1.45 & 0.19 & \phn7.8 & Ly$\alpha$ & 0.717 & Blend with line 20\\
22 & 2137.66 & 0.11 & 0.25 & 0.22 & 14.5 & SiII 1527 & 0.4 & Uncertain ID\\
\enddata
\end{deluxetable}

\clearpage

\begin{deluxetable}{ccccccccc}
\footnotesize
\tablecaption{Absorption Lines in LBQS 0107$-$025A (G190H)}
\label{tbl:table3}
\tablewidth{6in}
\tablewidth{0pt}
\tablehead{\colhead{No.} & \colhead{$\lambda$} &\colhead{$\sigma_{\lambda}$}
& \colhead{$W_{obs}$} & \colhead{$\sigma_{W}$} & \colhead{SNR} &\colhead{ID} &\colhead{$z$} &\colhead{Comments} \\ \colhead{} & \colhead{(\AA)} & \colhead{(\AA)} & \colhead{(\AA)} & \colhead{(\AA)}}
\startdata
1 & 1670.33 & 0.20 & 0.93 & 0.14 & \phn6.6 & SiII 1193& 0.4 & Blend w/ Galactic AlII 1670\\
2 & 1681.84 & 0.25 & 1.32 & 0.21 & \phn6.4 & Ly$\beta$ & 0.64 & Blend w/ SiII at $z$ = 0.29\\
3 & 1700.14 & 0.20 & 0.74 & 0.14 & \phn5.2 & Ly$\alpha$ & 0.399 & \\
4 & 1737.00 & 0.09 & 0.83 & 0.08 & 10.6 & Ly$\alpha$ & 0.429 & \\
5 & 1761.86 & 0.09 & 1.05 & 0.08 & 13.3 & Ly$\beta$ & 0.718 & \\
6 & 1772.37 & 0.11 & 0.47 & 0.07 & \phn6.7 & Ly$\beta$ & 0.728 & \\
7 & 1781.35 & 0.24 & 0.80 & 0.17 & \phn4.9 & Ly$\alpha$ & 0.465 & Weak line sample\\
8 & 1807.49 & 0.12 & 0.85 & 0.09 & \phn9.4 & SiIV 1402 & 0.29 & Blend w/ Galactic SiII 1807\\ 
9 & 1823.10 & 0.63 & 0.95 & 0.81 & \phn1.2 & OI 1302 & 0.4 & Blend with line 10\\
10 & 1824.52 & 0.39 & 1.04 & 0.80 & \phn1.3 & Ly$\alpha$ & 0.501 & \\
11 & 1831.57 & 0.14 & 0.81 & 0.09 & \phn9.1 & Ly$\beta$ & 0.786 \\
12 & 1866.15 & 0.12 & 2.36 & 0.26 & 27.3 & Ly$\alpha$ & 0.535 \\
13 & 1871.82 & 0.49 & 0.69 & 0.20 & \phn3.4 & Ly$\alpha$ & 0.540 & Weak line sample\\
14 & 1892.48 & 0.09 & 0.59 & 0.07 & \phn8.6 & Ly$\alpha$ & 0.557 & \\
15 & 1898.98 & 0.12 & 0.37 & 0.06 & \phn4.4 & Ly$\alpha$ & 0.562 & Weak line sample\\
16 & 1901.72 & 0.14 & 0.67 & 0.10 & \phn6.9 & Ly$\alpha$ & 0.564 & Blend with line 15\\
17 & 1910.68 & 0.07 & 0.47 & 0.04 & 11.3 & Ly$\alpha$ & 0.572 & \\
18 & 1923.80 & 0.08 & 0.59 & 0.07 & \phn9.1 & Ly$\beta$ & 0.876 & \\
19 & 1954.34 & 0.13 & 0.33 & 0.04 & \phn8.3 & Ly$\alpha$ & 0.608 & \\
20 & 1965.72 & 0.40 & 0.38 & 0.09 & \phn4.3 & Ly$\alpha$ & 0.617 & Weak line sample\\ 
21 & 1969.75 & 0.19 & 0.19 & 0.05 & \phn3.5 & Ly$\alpha$ & 0.620 & Weak line sample\\
22 & 1975.13 & 0.19 & 0.41 & 0.08 & \phn5.3 & Ly$\alpha$ & 0.625 & \\ 
23 & 1984.49 & 0.11 & 0.30 & 0.04 & \phn8.4 & Ly$\alpha$ & 0.632 & Blend with line 24\\ 
24 & 1986.34 & 0.11 & 0.26 & 0.04 & \phn6.1 & Ly$\alpha$ & 0.634 & \\
25 & 1993.21 & 0.06 & 0.48 & 0.03 & 15.1 & Ly$\alpha$ & 0.640 & \\
26 & 2004.29 & 0.49 & 0.36 & 0.10 & \phn3.7 & Ly$\alpha$ & 0.649 & Weak line sample\\
27 & 2006.19 & 0.08 & 0.39 & 0.07 & \phn5.3 & Ly$\alpha$ & 0.650 & Blend with line 26\\
28 & 2009.51 & 0.18 & 0.19 & 0.04 & \phn4.6 & Ly$\alpha$ & 0.653 & Weak line sample\\
29 & 2012.99 & 0.08 & 0.22 & 0.02 & \phn9.3 & Ly$\alpha$ & 0.656 & Blend with lines 30+31\\
30 & 2014.51 & 0.11 & 0.39 & 0.03 & 12.3 & Ly$\alpha$ & 0.657 & \\
31 & 2015.77 & 0.18 & 0.23 & 0.05 & \phn5.1 & Ly$\alpha$ & 0.658 & \\
32 & 2025.00 & 0.09 & 0.26 & 0.03 & \phn9.4 & Ly$\alpha$ & 0.666 & \\
33 & 2040.39 & 0.06 & 0.36 & 0.03 & 12.3 & Ly$\alpha$ & 0.678 & \\
34 & 2053.23 & 0.06 & 0.53 & 0.04 & 13.5 & Ly$\alpha$ & 0.689 & \\
35 & 2058.66 & 0.09 & 0.35 & 0.04 & \phn8.2 & Ly$\alpha$ & 0.693 & \\
36 & 2072.73 & 0.16 & 0.19 & 0.04 & \phn4.8 & FeII 1608 & 0.29 & \\
37 & 2088.53 & 0.14 & 1.21 & 0.24 & \phn5.1 & Ly$\alpha$ & 0.718 & \\
38 & 2100.40 & 0.04 & 1.23 & 0.05 & 25.0 & Ly$\alpha$ & 0.728 & \\
39 & 2121.99 & 0.19 & 0.25 & 0.05 & \phn4.6 & Ly$\alpha$ & 0.746 & Weak line sample\\
40 & 2166.33 & 0.13 & 0.22 & 0.04 & \phn5.8 & Ly$\alpha$ & 0.782 & \\ 
41 & 2170.81 & 0.04 & 1.29 & 0.05 & 25.8 & Ly$\alpha$ & 0.786 & \\
42 & 2195.95 & 0.10 & 0.21 & 0.03 & \phn7.0 & Ly$\alpha$ & 0.806 & \\
43 & 2240.10 & 0.14 & 0.40 & 0.05 & \phn8.0 & Ly$\alpha$ & 0.843 & \\
44 & 2245.40 & 0.04 & 0.52 & 0.03 & 16.1 & Ly$\alpha$ & 0.847 & \\ 
45 & 2250.50 & 0.08 & 0.37 & 0.04 & 10.4 & Ly$\alpha$ & 0.851 & \\
46 & 2261.52 & 0.39 & 0.46 & 0.09 & \phn5.2 & Ly$\alpha$ & 0.860 & \\
47 & 2270.81 & 0.06 & 0.40 & 0.03 & 12.7 & Ly$\alpha$ & 0.868 & \\
48 & 2280.33 & 0.02 & 1.09 & 0.03 & 38.3 & Ly$\alpha$ & 0.876 & \\
49 & 2296.08 & 0.05 & 0.46 & 0.03 & 14.0 & Ly$\alpha$ & 0.889 & \\
50 & 2307.66 & 0.07 & 0.42 & 0.03 & 13.3 & Ly$\alpha$ & 0.898 & \\  
\enddata
\end{deluxetable}

\clearpage

\begin{deluxetable}{ccccccccc}
\footnotesize
\tablecaption{Absorption Lines in LBQS 0107$-$025B (G190H)}
\label{tbl:table4}
\tablewidth{6in}
\tablewidth{0pt}
\tablehead{\colhead{No.} & \colhead{$\lambda$} &\colhead{$\sigma_{\lambda}$}
& \colhead{$W_{obs}$} & \colhead{$\sigma_{W}$} & \colhead{SNR} &\colhead{ID} &\colhead{$z$} &\colhead{Comments} \\ \colhead{} & \colhead{(\AA)} & \colhead{(\AA)} & \colhead{(\AA)} & \colhead{(\AA)}}
\startdata
1 & 1669.90 & 0.13 & 0.89 & 0.15 & \phn6.0 & AlII 1670 & 0.001 & \\
2 & 1699.80 & 0.07 & 1.31 & 0.12 & 11.2 & Ly$\alpha$ & 0.398 & \\
3 & 1716.33 & 0.33 & 1.05 & 0.20 & \phn5.2 & Ly$\alpha$ & 0.412 & \\
4 & 1736.81 & 0.18 & 0.50 & 0.11 & \phn4.4 & Ly$\gamma$ & 0.786 & \\
5 & 1745.28 & 0.11 & 0.63 & 0.11 & \phn6.2 & Ly$\xi$  & 0.876 & \\
6 & 1758.65 & 0.14 & 0.52 & 0.12 & \phn4.5 & Ly$\gamma$ & 0.808 & \\
7 & 1761.33 & 0.09 & 1.58 & 0.16 & 10.2 & Ly$\beta$ & 0.717 & \\
8 & 1781.00 & 0.21 & 0.48 & 0.13 & \phn3.7 & Ly$\delta$ & 0.875 & \\
9 & 1783.35 & 0.43 & 0.78 & 0.21 & \phn3.7 & Ly$\alpha$ & 0.467 & Weak line sample\\
10 & 1822.47 & 0.09 & 1.72 & 0.11 & 15.7 & Ly$\alpha$ & 0.499 & \\
11 & 1832.13 & 0.19 & 0.36 & 0.09 & \phn4.0 & Ly$\beta$ & 0.786 & \\
12 & 1843.14 & 0.07 & 0.78 & 0.08 & 10.3 & Ly$\alpha$ & 0.516 & \\
13 & 1851.50 & 0.12 & 0.49 & 0.09 & \phn5.4 & Ly$\alpha$ & 0.523 & \\
14 & 1854.48 & 0.10 & 1.20 & 0.11 & 11.0 & Ly$\beta$ & 0.808 & \\
15 & 1865.58 & 0.10 & 1.05 & 0.11 & 10.0 & Ly$\alpha$ & 0.535 & \\
16 & 1890.14 & 0.23 & 0.32 & 0.09 & \phn3.5 & Ly$\alpha$ & 0.555 & Weak line sample\\
17 & 1918.12 & 0.24 & 0.62 & 0.11 & \phn5.6 & Ly$\alpha$ & 0.578 & \\
18 & 1923.56 & 0.08 & 0.97 & 0.09 & \phn5.4 & Ly$\beta$ & 0.875 & \\
19 & 1949.19 & 0.19 & 0.29 & 0.08 & \phn3.8 & SiIV 1393 & 0.398 & \\
20 & 1955.30 & 0.21 & 0.47 & 0.10 & \phn4.7 & Ly$\alpha$ & 0.608 & Weak line sample\\
21 & 1970.91 & 0.15 & 0.37 & 0.07 & \phn5.0 & Ly$\alpha$ & 0.621 & \\
22 & 1996.16 & 0.28 & 0.49 & 0.11 & \phn4.7 & Ly$\alpha$ & 0.642 & Weak line sample\\ 
23 & 2000.99 & 0.28 & 0.30 & 0.08 & \phn3.7 & Ly$\alpha$ & 0.646 & Weak line sample\\
24 & 2004.80 & 0.31 & 0.33 & 0.10 & \phn3.5 & Ly$\alpha$ & 0.649 & Weak line sample\\
25 & 2017.43 & 0.16 & 0.33 & 0.06 & \phn5.5 & Ly$\alpha$ & 0.660 & \\
26 & 2047.72 & 0.32 & 0.25 & 0.07 & \phn3.5 & Ly$\alpha$ & 0.684 & Weak line sample\\
27 & 2053.52 & 0.12 & 0.34 & 0.05 & \phn6.2 & Ly$\alpha$ & 0.689 & \\
28 & 2082.58 & 0.10 & 0.55 & 0.06 & \phn8.9 & Ly$\alpha$ & 0.713 & \\
29 & 2087.83 & 0.03 & 1.37 & 0.05 & 27.6 & Ly$\alpha$ & 0.717 & \\
30 & 2117.35 & 0.29 & 0.23 & 0.07 & \phn3.6 & Ly$\alpha$ & 0.742 & Weak line sample\\
31 & 2124.43 & 0.16 & 0.30 & 0.06 & \phn4.9 & Ly$\alpha$ & 0.748 & Weak line sample\\
32 & 2165.25 & 0.12 & 0.45 & 0.06 & \phn7.4 & CIV 1548 & 0.398 & \\
33 & 2169.07 & 0.15 & 0.38 & 0.07 & \phn5.4 & CIV 1550 & 0.398 & \\
34 & 2171.85 & 0.06 & 0.78 & 0.07 & 10.9 & Ly$\alpha$ & 0.787 & \\
35 & 2184.49 & 0.16 & 0.53 & 0.07 & \phn7.8 & Ly$\alpha$ & 0.797 & \\
36 & 2198.64 & 0.03 & 1.61 & 0.05 & 31.2 & Ly$\alpha$ & 0.809 & \\
37 & 2209.69 & 0.11 & 0.30 & 0.05 & \phn6.1 & Ly$\alpha$ & 0.818 & \\
38 & 2214.66 & 0.49 & 0.40 & 0.10 & \phn4.3 & Ly$\alpha$ & 0.822 & Weak line sample\\
39 & 2225.93 & 0.08 & 0.48 & 0.05 & \phn9.3 & Ly$\alpha$ & 0.831 & \\
40 & 2229.33 & 0.12 & 0.38 & 0.05 & \phn6.9 & Ly$\alpha$ & 0.834 & \\ 
41 & 2245.64 & 0.22 & 0.27 & 0.06 & \phn4.5 & Ly$\alpha$ & 0.847 & Weak line sample\\
42 & 2279.89 & 0.03 & 1.10 & 0.04 & 25.2 & Ly$\alpha$ & 0.875 & \\
43 & 2296.54 & 0.09 & 0.27 & 0.04 & \phn6.6 & Ly$\alpha$ & 0.889 & \\
\enddata
\tablecomments{This data was taken from \citet{dinth97}. Plots of the spectrum are available there. A wavelength offset of -1.72 \AA\ was applied to remove a discrepancy in the wavelength zero point determination between the two data sets.}
\end{deluxetable}

\clearpage

\begin{deluxetable}{ccccccccc}
\footnotesize
\tablecaption{Absorption Lines in LBQS 0107$-$025A (G270H)}
\label{tbl:table5}
\tablewidth{6in}
\tablewidth{0pt}
\tablehead{\colhead{No.} & \colhead{$\lambda$} &\colhead{$\sigma_{\lambda}$}
& \colhead{$W_{obs}$} & \colhead{$\sigma_{W}$} & \colhead{SNR} &\colhead{ID} &\colhead{$z$} &\colhead{Comments} \\ \colhead{} & \colhead{(\AA)} & \colhead{(\AA)} & \colhead{(\AA)} & \colhead{(\AA)}}
\startdata
1 & 2245.74 & 0.08 & 0.39 & 0.03 & \phn14.6 & Ly$\alpha$ & 0.847 & \\
2 & 2250.39 & 0.14 & 0.43 & 0.03 & \phn14.5 & FeII 1608& 0.4& Blend w/ Galactic FeII 2249\\
3 & 2270.81 & 0.11 & 0.39 & 0.03 & \phn15.3 & Ly$\alpha$ & 0.868 & \\
4 & 2280.11 & 0.04 & 1.40 & 0.06 & \phn25.2 & Ly$\alpha$ & 0.876 & \\
5 & 2295.94 & 0.09 & 0.35 & 0.02 & \phn15.7 & Ly$\alpha$ & 0.889 & \\
6 & 2307.40 & 0.12 & 0.44 & 0.04 & \phn10.4 & Ly$\alpha$ & 0.898 & Excluded by proximity effect \\
7 & 2316.67 & 0.05 & 1.08 & 0.04 & \phn25.1 & Ly$\alpha$ & 0.906 & Excluded by proximity effect\\
8 & 2340.91 & 0.05 & 2.32 & 0.04 & \phn61.9 & Ly$\alpha$ & 0.926 & Excluded by proximity effect\\
9 & 2359.86 & 0.02 & 0.58 & 0.02 & \phn34.9 & Ly$\alpha$ & 0.941 & Excluded by proximity effect\\
10 & 2369.18 & 0.07 & 0.42 & 0.10 & \phn\phn4.2 & Ly$\alpha$ & 0.949 & Excluded by proximity effect\\
11 & 2373.20 & 0.08 & 0.25 & 0.01 & \phn27.1 & FeII 2374 & 0.0 & \\ 
12 & 2377.34 & 0.03 & 0.58 & 0.02 & \phn25.9 & Ly$\alpha$ & 0.956 & Excluded by proximity effect\\
13 & 2381.70 & 0.04 & 0.28 & 0.01 & \phn19.6 & FeII 2382 & 0.0 \\
14 & 2388.01 & 0.05 & 0.45 & 0.02 & \phn29.1 & ... & ... & \\
15 & 2394.88 & 0.23 & 0.08 & 0.02 & \phn\phn4.5 & ... & ... & \\
16 & 2449.24 & 0.22 & 0.13 & 0.03 & \phn\phn4.3 & ... & ... & \\
17 & 2585.51 & 0.11 & 0.61 & 0.05 & \phn11.6 & FeII 2586 & 0.0 & \\
18 & 2598.76 & 0.03 & 0.61 & 0.01 & 101.8 & FeII 2600 & 0.0 & \\
19 & 2614.14 & 0.11 & 0.28 & 0.03 & \phn\phn8.4 & MnII 2606 & 0.003 & \\
20 & 2700.44 & 0.28 & 0.18 & 0.06 & \phn\phn3.2 & ... & ... & \\
21 & 2794.88 & 0.04 & 1.00 & 0.03 & \phn33.6 & ... & ... & \\
22 & 2802.07 & 0.05 & 0.82 & 0.04 & \phn20.7 & ... & ... & \\
23 & 2851.59 & 0.09 & 0.27 & 0.02 & \phn12.4 & ... & ... & \\
24 & 2903.73 & 0.12 & 0.34 & 0.05 & \phn\phn6.8 & ... & ... & \\
25 & 2922.74 & 0.28 & 0.15 & 0.04 & \phn\phn3.5 & FeII 2382 & 0.228 & \\
\enddata
\end{deluxetable}

\clearpage

\begin{deluxetable}{ccccccccc}
\footnotesize
\tablecaption{Absorption Lines in LBQS 0107$-$025B (G270H)}
\label{tbl:table6}
\tablewidth{6in}
\tablewidth{0pt}
\tablehead{\colhead{No.} & \colhead{$\lambda$} &\colhead{$\sigma_{\lambda}$}
& \colhead{$W_{obs}$} & \colhead{$\sigma_{W}$} & \colhead{SNR} &\colhead{ID} &\colhead{$z$} &\colhead{Comments} \\ \colhead{} & \colhead{(\AA)} & \colhead{(\AA)} & \colhead{(\AA)} & \colhead{(\AA)}}
\startdata
1 & 2226.21 & 0.25 & 0.16 & 0.05 & \phn\phn3.4 & Ly$\alpha$ & 0.831 & Weak line sample \\
2 & 2245.32 & 0.13 & 0.47 & 0.04 & \phn11.7 & Ly$\alpha$ & 0.847 & \\
3 & 2279.81 & 0.03 & 1.44 & 0.03 & \phn45.3 & Ly$\alpha$ & 0.875 & \\
4 & 2296.68 & 0.17 & 0.19 & 0.03 & \phn\phn6.0 & Ly$\alpha$ & 0.889 & \\
5 & 2307.39 & 0.21 & 0.30 & 0.04 & \phn\phn8.2 & Ly$\alpha$ & 0.898 & Excluded by proximity effect\\
6 & 2317.70 & 0.02 & 1.04 & 0.02 & \phn56.9 & Ly$\alpha$ & 0.907 & Excluded by proximity effect\\
7 & 2340.17 & 0.06 & 0.33 & 0.03 & \phn11.1 & AlII 1670 & 0.4 & \\
8 & 2343.20 & 0.05 & 1.03 & 0.05 & \phn21.2 & FeII 2344 & 0.0 & \\
9 & 2347.84 & 0.05 & 0.25 & 0.02 & \phn15.4 & Ly$\alpha$ & 0.931 & Excluded by proximity effect\\
10 & 2353.52 & 0.05 & 0.54 & 0.02 & \phn29.3 & Ly$\alpha$ & 0.936 & Excluded by proximity effect\\
11 & 2360.32 & 0.09 & 0.09 & 0.01 & \phn\phn9.7 & Ly$\alpha$ & 0.942 & Excluded by proximity effect\\
12 & 2373.76 & 0.08 & 0.20 & 0.00 & 129.7 & FeII 2374 & 0.0 & Excluded by proximity effect\\
13 & 2381.88 & 0.02 & 0.60 & 0.01 & \phn59.3 & FeII 2382 & 0.0 & \\
14 & 2576.17 & 0.14 & 0.20 & 0.03 & \phn\phn6.3 & MnII 2576 & 0.0 & \\
15 & 2585.87 & 0.04 & 0.50 & 0.02 & \phn33.4 & FeII 2586 & 0.0 & \\
16 & 2599.26 & 0.06 & 0.69 & 0.03 & \phn24.5 & FeII 2599 & 0.0 & \\
17 & 2795.17 & 0.01 & 1.05 & 0.02 & \phn61.4 & MgII 2796 & 0.0 & \\
18 & 2802.49 & 0.04 & 1.00 & 0.03 & \phn33.3 & MgII 2803 & 0.0 & \\
19 & 2852.34 & 0.14 & 0.31 & 0.04 & \phn\phn7.1 & MgI 2852 & 0.0 & \\
20 & 3017.10 & 0.16 & 0.15 & 0.01 & \phn23.3 & ... & ... & \\ 
21 & 3173.58 & 0.21 & 0.27 & 0.04 & \phn\phn6.9 & FeII 2586 & 0.227 & \\
22 & 3177.95 & 0.33 & 0.25 & 0.05 & \phn\phn5.0 & ... & ... \\ 
\enddata
\end{deluxetable}
        
\clearpage

\begin{deluxetable}{cccccccc}
\footnotesize
\tablecaption{Matches in the Pair LBQS 0107$-$0232 and LBQS 0107$-$025A}
\label{tbl:table7}
\tablewidth{6in}
\tablewidth{0pt}
\tablehead{\colhead{No.} & \colhead{Line No.} & \colhead{Line No.}& \colhead{$\lambda_{0232}$} &\colhead{$\lambda_{025A}$} & \colhead{$W_{obs(0232)}$} & \colhead{$W_{obs(025A)}$} & \colhead{$|\Delta$v$|$} \\ \colhead{} &\colhead{0232} & \colhead{025A} & \multicolumn{2}{c}{(\AA)} & \multicolumn{2}{c}{(\AA)} & \colhead{$\rm{(km\ s^{-1})}$}}
\startdata
1 & 7 & 12 & 1864.19 & 1866.15 & 2.48 & 2.36 & 264 \\
2 & 9 & 14 & 1892.33 & 1892.48 & 3.63 & 0.59 & \phn20 \\
3 & 11 & 19 & 1961.90 & 1954.34 & 1.09 & 0.33 & 930 \\
4 & 12 & 22 & 1970.57 & 1975.13 & 0.64 & 0.41 & 553 \\
5 & 13 & 27 & 2003.37 & 2006.19 & 0.68 & 0.39 & 332 \\
6 & 14 & 32 & 2026.79 & 2025.00 & 1.08 & 0.26 & 206 \\
7 & 15 & 33 & 2045.57 & 2040.39 & 0.70 & 0.36 & 587 \\
8 & 16 & 34 & 2053.37 & 2053.23 & 1.00 & 0.53 & \phn16 \\
9 & 21 & 37 & 2087.81 & 2088.53 & 1.45 & 1.21 & \phn78 \\
\enddata
\end{deluxetable}

\begin{deluxetable}{cccccccc}
\footnotesize
\tablecaption{Matches in the Pair LBQS 0107$-$0232 and LBQS 0107$-$025B}
\label{tbl:table8}
\tablewidth{6in}
\tablewidth{0pt}
\tablehead{\colhead{No.} & \colhead{Line No.} & \colhead{Line No.}& \colhead{$\lambda_{0232}$} &\colhead{$\lambda_{025B}$}& \colhead{$W_{obs(0232)}$} & \colhead{$W_{obs(025B)}$} & \colhead{$|\Delta$v$|$} \\ \colhead{} &\colhead{0232} &\colhead{025B} & \multicolumn{2}{c}{(\AA)} & \multicolumn{2}{c}{(\AA)}& \colhead{$\rm{(km\ s^{-1})}$}}
\startdata
1 & 5 & 13 & 1853.27 & 1851.50 & 0.87 & 0.49 & \phn241 \\
2 & 7 & 15 & 1864.19 & 1865.58 & 2.48 & 1.05 & \phn187 \\
3 & 10 & 17 & 1918.21 & 1918.12 & 0.70 & 0.62 & \phn\phn11 \\
4 & 12 & 21 & 1970.57 & 1970.91 & 0.64 & 0.37 & \phn\phn41 \\
5 & 14 & 25 & 2026.79 & 2017.43 & 1.08 & 0.33 & 1083 \\
6 & 16 & 27 & 2053.37 & 2053.52 & 1.00 & 0.34 & \phn\phn17 \\
7 & 19 & 28 & 2079.21 & 2082.58 & 0.42 & 0.55 & \phn369 \\
8 & 21 & 29 & 2087.81 & 2087.83 & 1.45 & 1.37 & \phn\phn\phn2 \\
\enddata
\end{deluxetable}

\begin{deluxetable}{cccccccc}
\footnotesize
\tablecaption{Matches in the Pair LBQS 0107$-$025A and LBQS 0107$-$025B}
\label{tbl:table9}
\tablewidth{6in}
\tablewidth{0pt}
\tablehead{\colhead{No.} & \colhead{Line No.} & \colhead{Line No.}& \colhead{$\lambda_{025A}$} &\colhead{$\lambda_{025B}$}& \colhead{$W_{obs(025A)}$} & \colhead{$W_{obs(025B)}$} & \colhead{$|\Delta$v$|$} \\ \colhead{} &\colhead{025A} &\colhead{025B} & \multicolumn{2}{c}{(\AA)}& \multicolumn{2}{c}{(\AA)} & \colhead{$\rm{(km\ s^{-1})}$}}
\startdata
1 & 3 & 2 & 1700.14 & 1699.80 & 0.74 & 1.31 & \phn\phn54 \\
2 & 12 & 15 & 1866.15 & 1865.58 & 2.36 & 1.05 & \phn\phn77 \\
3 & 17 & 17 & 1910.68 & 1918.12 & 0.47 & 0.62 & \phn955 \\
4 & 22 & 21 & 1975.13 & 1970.91 & 0.41 & 0.37 & \phn512 \\
5 & 31 & 25 & 2015.77 & 2017.43 & 0.23 & 0.33 & \phn193 \\
6 & 34 & 27 & 2053.23 & 2053.52 & 0.53 & 0.34 & \phn\phn32 \\
7 & 37 & 29 & 2088.53 & 2087.83 & 1.21 & 1.37 & \phn\phn76 \\
8 & 41 & 34 & 2170.81 & 2171.85 & 1.29 & 0.78 & \phn104 \\
9 & 42 & 36 & 2195.95 & 2198.64 & 0.21 & 1.61 & \phn264 \\
10 & 43 & 40 & 2240.10 & 2229.33 & 0.40 & 0.38 & 1019 \\
11 & 48 & 42 & 2280.33 & 2279.89 & 1.09 & 1.10 & \phn\phn40 \\
12 & 49 & 43 & 2296.08 & 2296.54 & 0.46 & 0.27 & \phn\phn41 \\
\enddata
\end{deluxetable}

\begin{deluxetable}{cccccccccccc}
\footnotesize
\tablecaption{Matches Common to all Three Quasars}
\label{tbl:table10}
\tablewidth{6in}
\tablewidth{0pt}
\tablehead{\multicolumn{3}{c}{LBQS 0107$-$0232} & \multicolumn{3}{c}{LBQS 0107$-$025A} & \multicolumn{3}{c}{LBQS 0107$-$025B} & \multicolumn{3}{c}{$|\Delta$v$|$} \\ \colhead{No.} &\colhead{$\lambda$} &\colhead{$W_{obs}$} & \colhead{No.} &\colhead{$\lambda$} & \colhead{$W_{obs}$} & \colhead{No.} & \colhead{$\lambda$} & \colhead{$W_{obs}$} \\ \colhead{} & \multicolumn{2}{c}{(\AA)} & \colhead{} & \multicolumn{2}{c}{(\AA)} & \colhead{} & \multicolumn{2}{c}{(\AA)} & \colhead{} & \colhead{$\rm{(km\ s^{-1})}$}}
\startdata
8 & 1864.19 & 2.48 & 12 & 1866.15 & 2.36 & 15 & 1865.58 & 1.05 & 264 & 418 & 77 \\
13 & 1970.57 & 0.64 & 22 & 1975.13 & 0.41 & 21 & 1970.91 & 0.37 & 533 & 250 & 512 \\
17 & 2053.37 & 1.00 & 35 & 2053.52 & 0.53 & 27 & 2053.52 & 0.34 & \phn16 & 209 & \phn32 \\
\enddata
\end{deluxetable}

\begin{deluxetable}{ccccc}
\footnotesize
\tablecaption{Probability of Observed Coincidences Occurring Randomly}
\label{tbl:table11}
\tablewidth{6in}
\tablewidth{0pt}
\tablehead{\colhead{$|\Delta$v$|$} & \colhead{LBQS 0107$-$0232,} & \colhead{LBQS 0107$-$0232,} & \colhead{LBQS 0107$-$025A,} & \colhead{Triple} \\ \colhead{($\rm{km\ s^{-1}}$)} & \colhead{LBQS 0107$-$025A} & \colhead{LBQS 0107$-$025B} & \colhead{LBQS 0107$-$025B}} 
\startdata
$<$ 200 & 0.5143 & 0.0989 & 0.0921 & 0.6922 \\
$<$ 400 & 0.3476 & 0.1137 & 0.1878 & 0.5776 \\
$<$ 600 & 0.3010 & 0.2011 & 0.2451 & 0.4666 \\
\enddata
\end{deluxetable}

\begin{figure}
\includegraphics[angle=-90,width=\textwidth,bb=102 16 476 784,clip=]{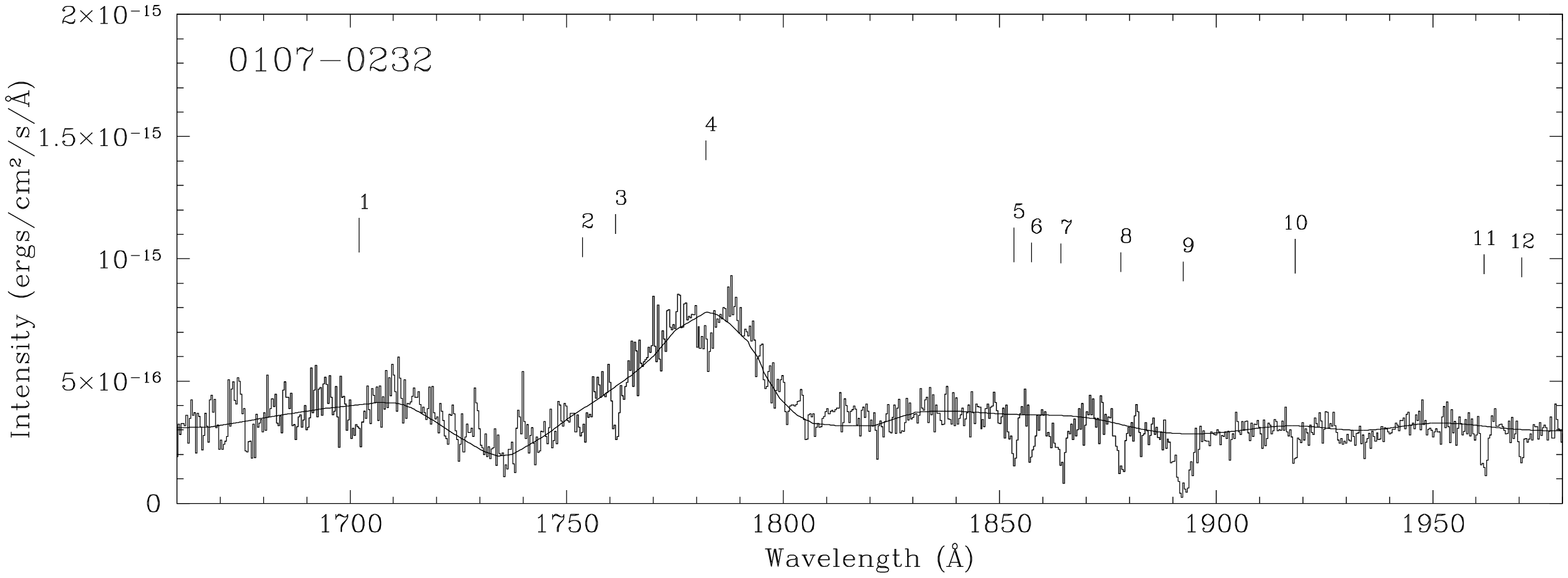}
\includegraphics[angle=-90,width=\textwidth,bb=102 16 476 784,clip=]{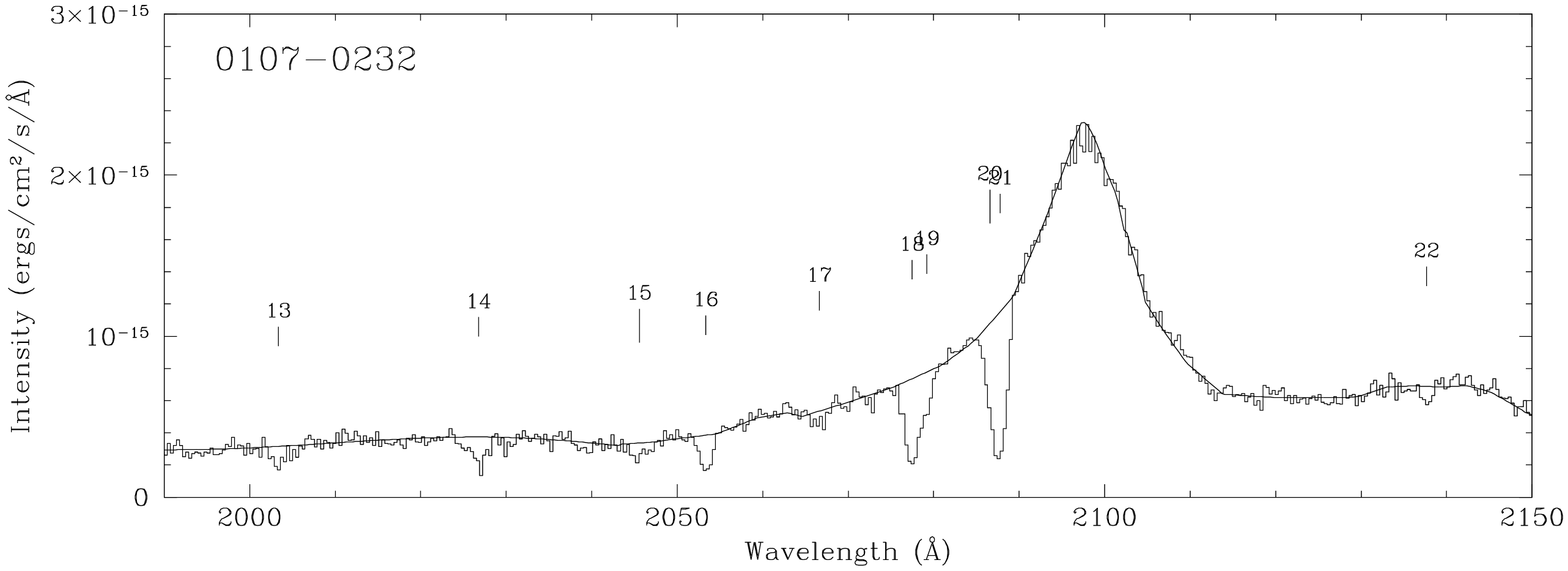}
\caption{Spectrum of LBQS 0107$-$0232 in G190H with identified lines marked}
\end{figure}

\clearpage

\begin{figure}
\includegraphics[angle=-90,width=0.75\textwidth,bb=102 16 476 784,clip=]{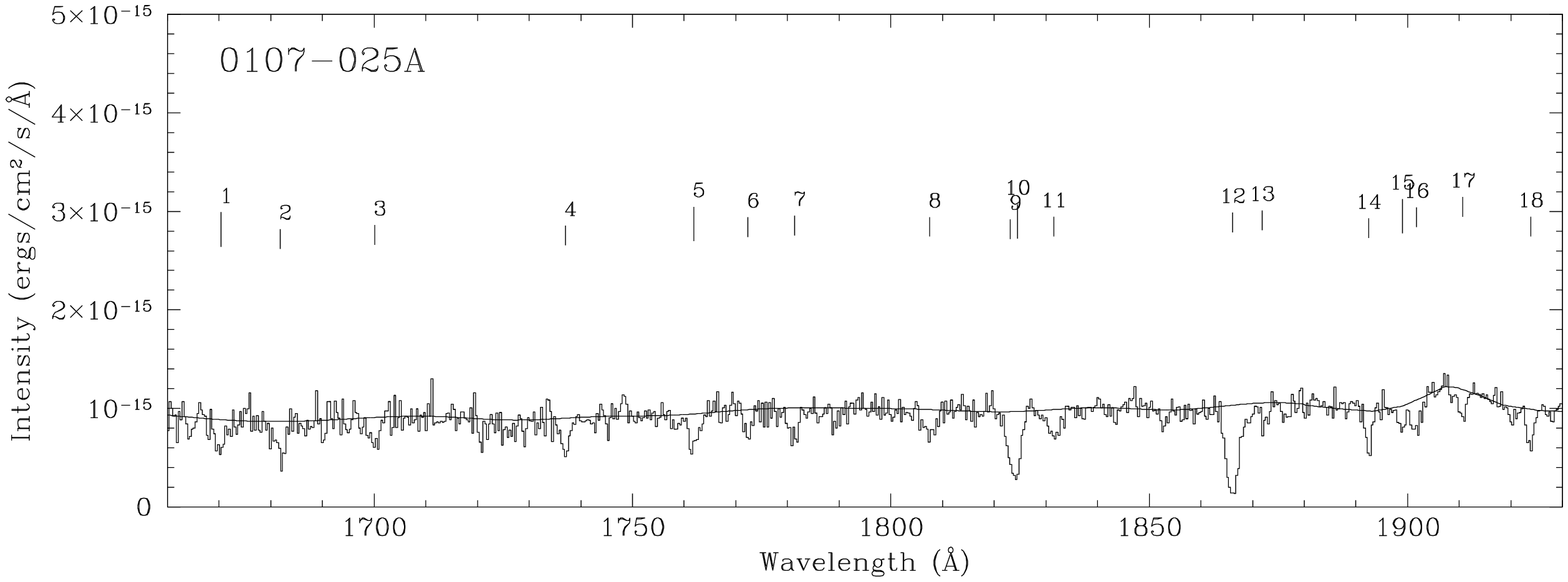}
\includegraphics[angle=-90,width=0.75\textwidth,bb=102 16 476 784,clip=]{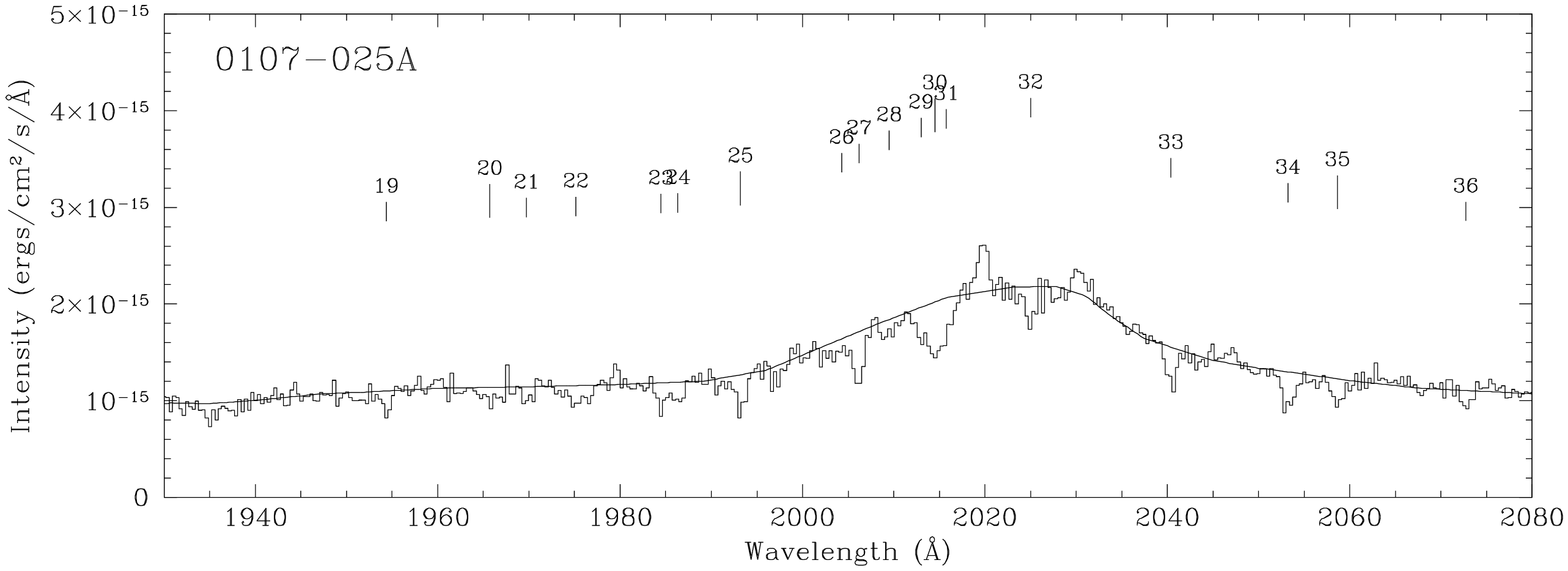}
\includegraphics[angle=-90,width=0.75\textwidth,bb=102 16 476 784,clip=]{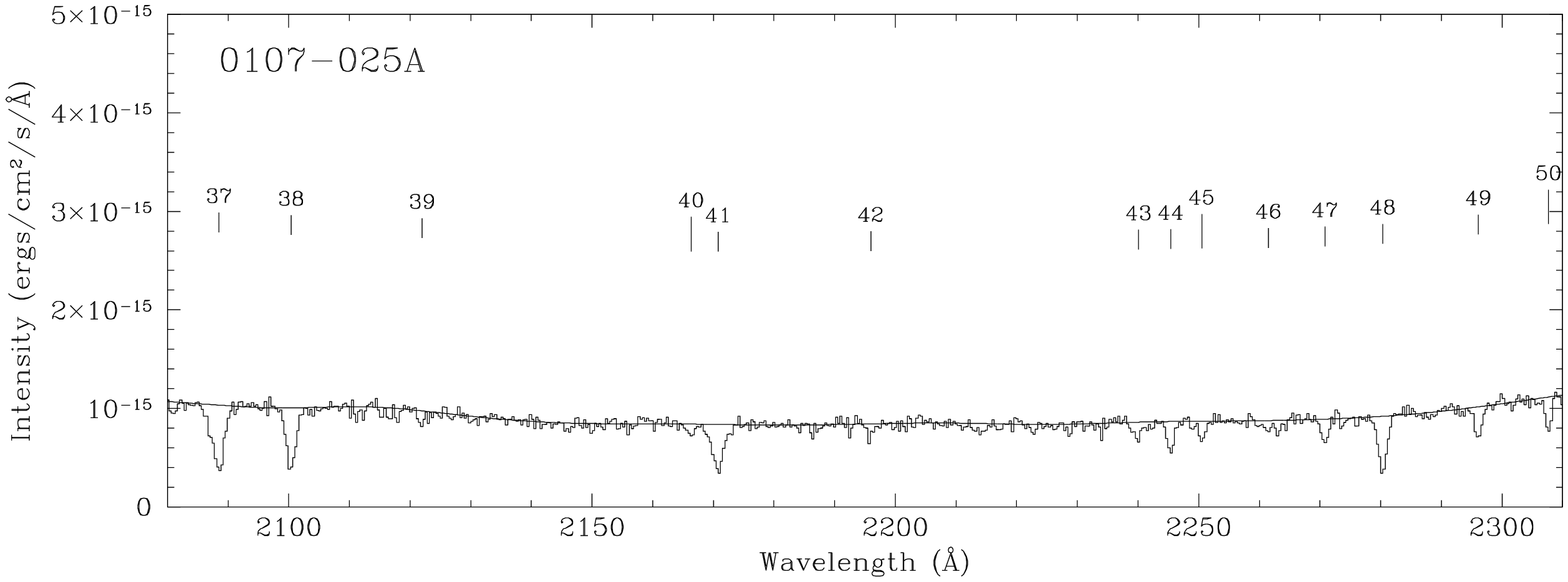}
\caption{Spectrum of LBQS 0107$-$025A in G190H with identified lines marked}
\end{figure}

\clearpage

\begin{figure}
\includegraphics[angle=-90,width=\textwidth,bb=102 16 476 784,clip=]{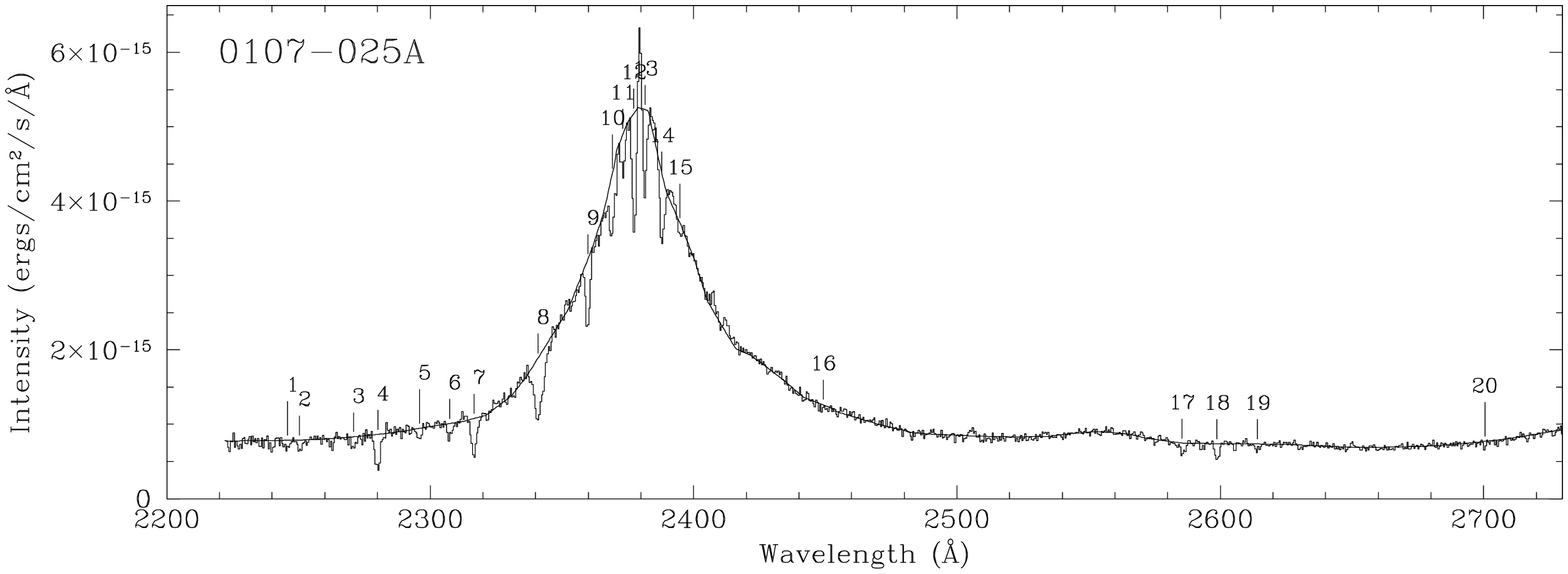}
\includegraphics[angle=-90,width=\textwidth,bb=102 16 476 784,clip=]{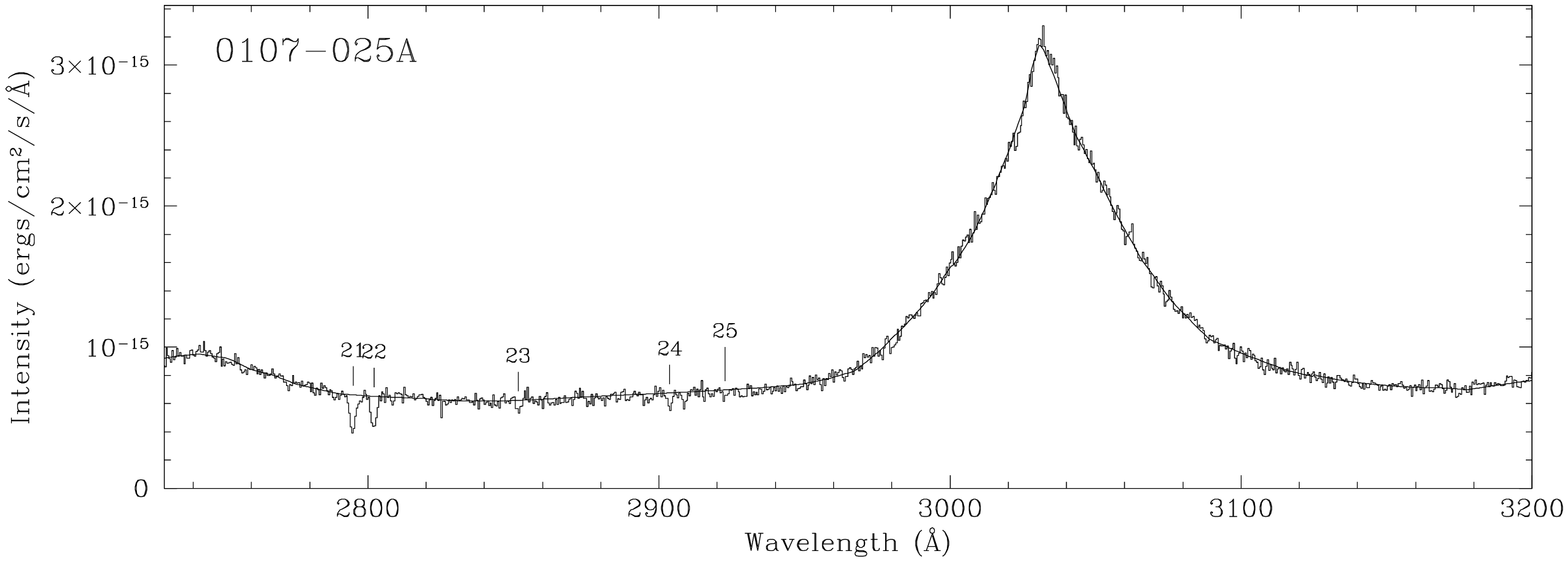}
\caption{Spectrum of LBQS 0107$-$025A in G270H with identified lines marked}
\end{figure}

\clearpage

\begin{figure}
\includegraphics[angle=-90,width=\textwidth,bb=102 16 476 784,clip=]{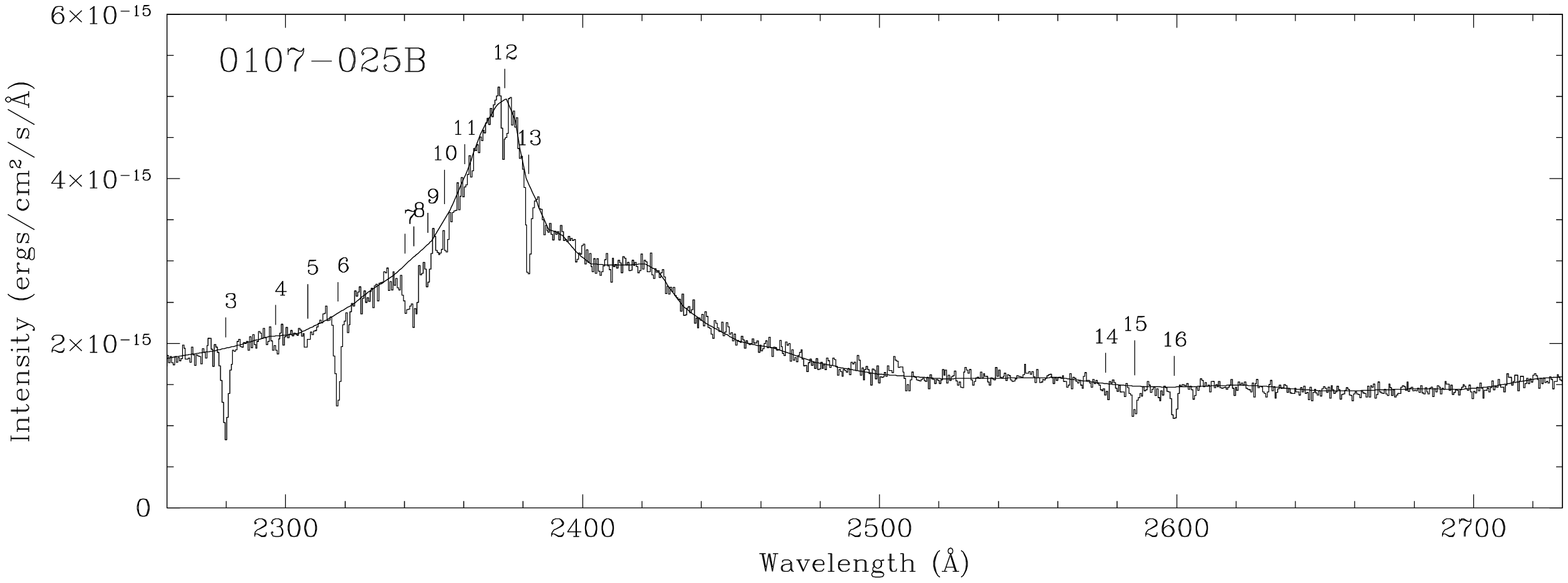}
\includegraphics[angle=-90,width=\textwidth,bb=102 16 476 784,clip=]{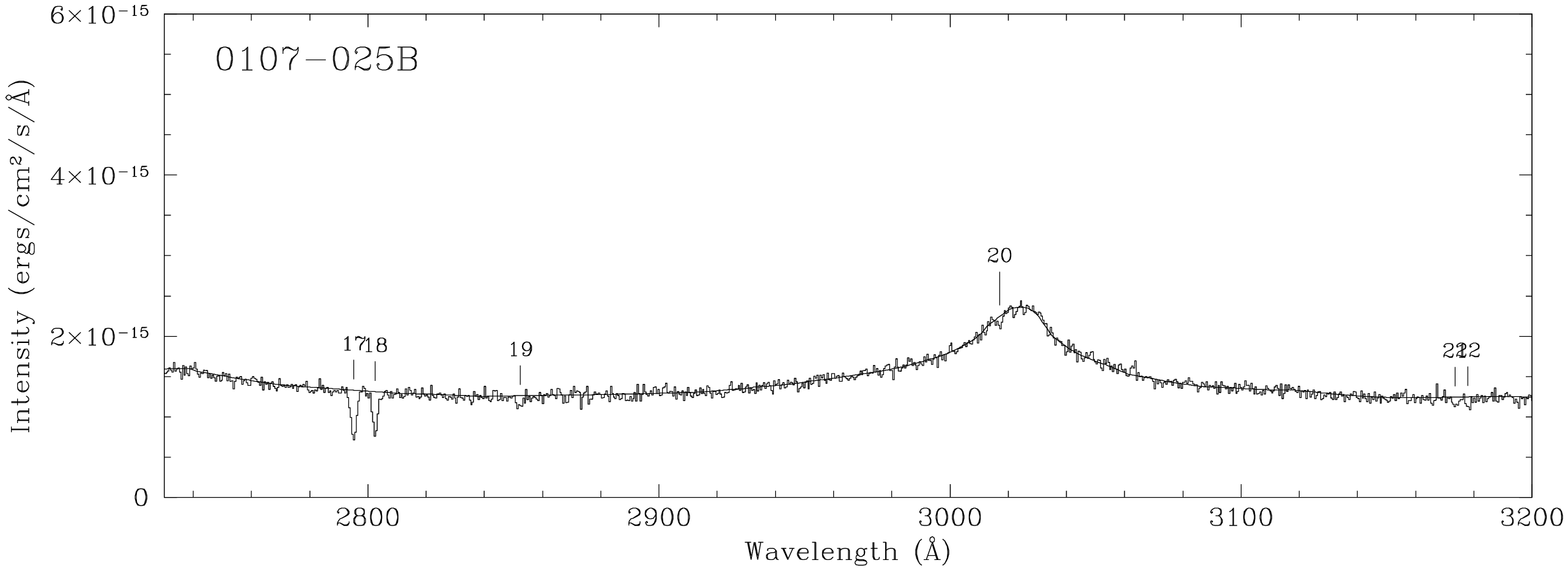}
\caption{Spectrum of LBQS 0107$-$025B in G270H with identified lines marked}
\end{figure}

\clearpage

\begin{figure}
\plotone{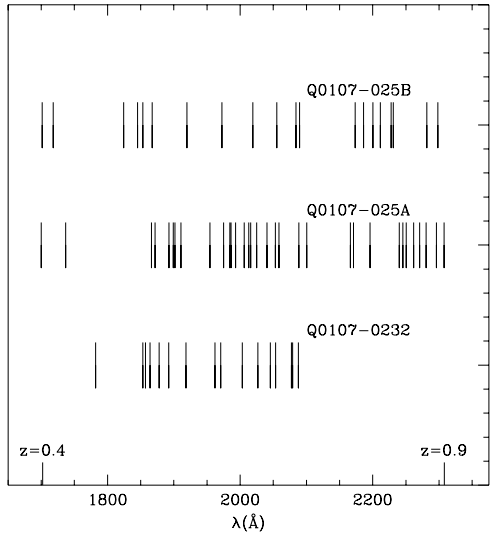}
\caption{Lines stronger than 5$\sigma$\ that were used in matching plotted side-by-side for comparison}
\end{figure}

\begin{figure}
\plotone{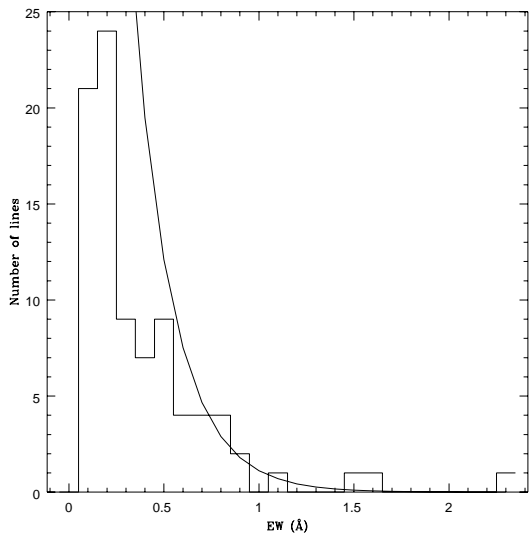}
\caption{Distribution of rest equivalent widths for the 3$\sigma$ line lists of all three LOSs binned to 0.1 \AA\ with the HST Key Project distribution from \citet{wey98} superimposed. The sample is substantially incomplete below 0.5\AA.}
\end{figure}

\begin{figure}
\plotone{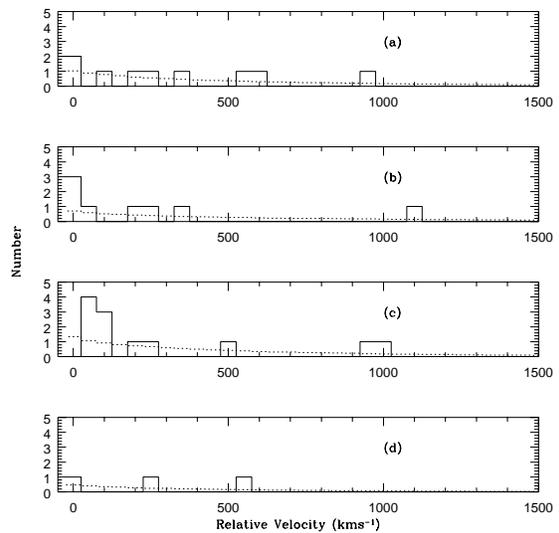}
\caption{Number of matches against $|\Delta$v$|$ for the pairings (a) LBQS 0107$-$0232,025A, (b) LBQS 0107$-$0232,025B, (c) LBQS 0107$-$025A,B, and (d) all three sight-lines taken together. Dotted lines are the expected number of matches from the corresponding Monte Carlo simulations. Velocity separations are binned to 50 $\rm{km\ s^{-1}}$ and each simulation is comprised of 10,000 trials.}
\end{figure}

\begin{figure}
\plotone{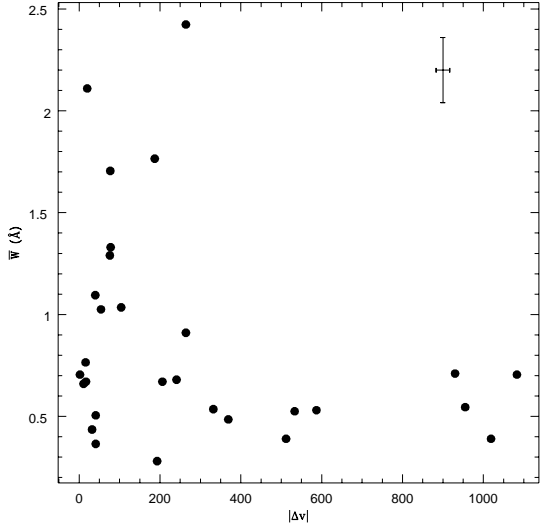}
\caption{Average observed equivalent width of each pair plotted against $|\Delta$v$|$. The error bar corresponds to the largest error in the sample. The strong lines preferentially have small velocity separations.}
\end{figure}

\begin{figure}
\plottwo{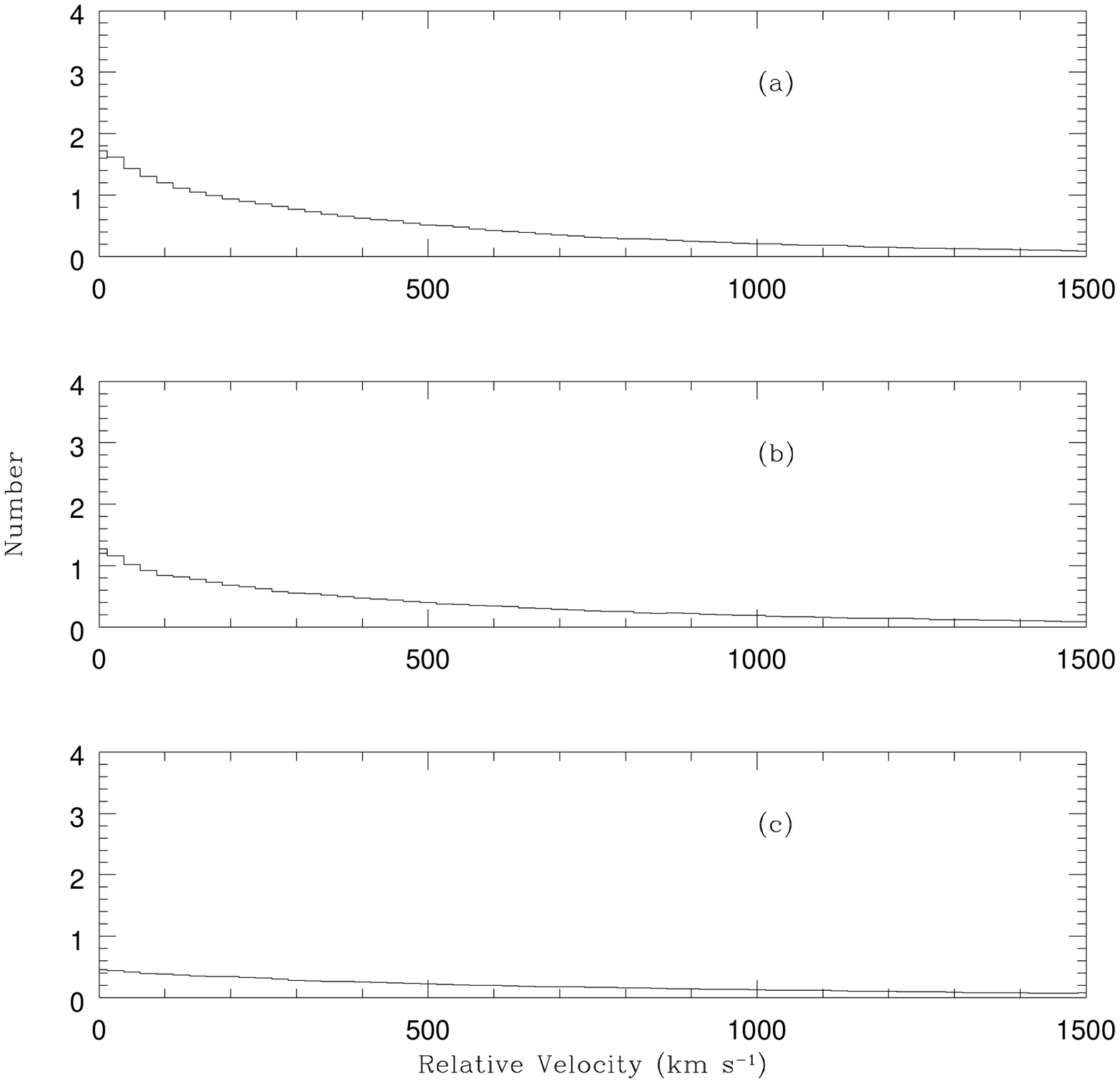}{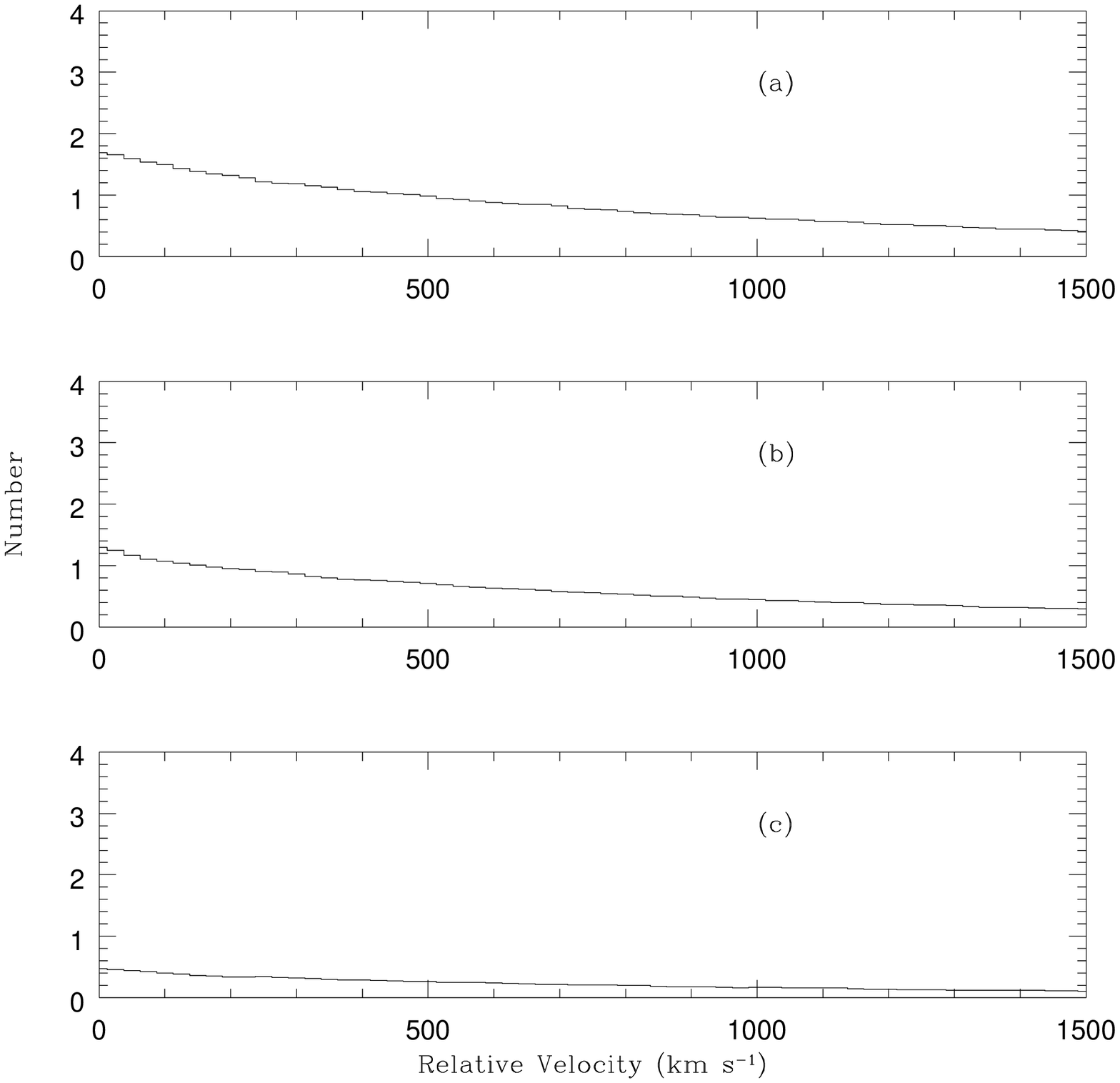}
\caption{Expected number of matches against $|\Delta$v$|$ from Monte Carlo simulations. The left panel shows matches with the uniqueness criterion in place for pairings with (a) 30 and 30, (b) 30 and 20, and (c) 30 and 10 lines in the respective lines of sight. The right panel shows the same set with the uniqueness criterion off. The histograms are binned to 25 $\rm{km\ s^{-1}}$, and there are ten thousand trials in each simulation.}
\end{figure}

\end{document}